\documentclass[a4paper,11pt]{article}
\pdfoutput=1 

\usepackage{jheppub} 

\usepackage{amssymb} 
\usepackage{amsmath}
\usepackage{mathtools}
\usepackage{amsfonts}    
\usepackage{dsfont}
\usepackage{pdfpages}
\usepackage{verbatim}
\usepackage{tensor}
\usepackage{mathrsfs}
\usepackage[mathscr]{euscript}
\usepackage{tikz}
\usepackage{braket}
\usepackage{physics}

\let\oldbibliography\thebibliography
\renewcommand{\thebibliography}[1]{\oldbibliography{#1}
\setlength{\itemsep}{4.14pt}} 

\allowdisplaybreaks
\usepackage{mathrsfs}
\usepackage[mathscr]{euscript}

\numberwithin{equation}{section}

\def\be{\begin{equation}}
\def\ee{\end{equation}}

\title{Stress-energy tensor correlators from the world-sheet}

\author{Hanno Bertle}
\author{\!\!, Andrea Dei}
\author{and Matthias R.~Gaberdiel}

\affiliation{Institut f\"ur Theoretische Physik, ETH Z\"urich \\
\hspace*{0.3cm} Wolfgang-Pauli-Stra{\ss}e 27, 8093 Z\"urich, Switzerland}

\emailAdd{bertleh@student.ethz.ch}
\emailAdd{adei@itp.phys.ethz.ch}
\emailAdd{gaberdiel@itp.phys.ethz.ch}

\abstract{The large $N$ limit of symmetric orbifold theories was recently argued to have an AdS/CFT dual world-sheet description in terms of an $\mathfrak{sl}(2,\mathds{R})$ WZW model. In previous work the world-sheet state corresponding to the symmetric orbifold stress-energy tensor was identified. We calculate certain 2- and 3-point functions of the corresponding vertex operator on the world-sheet, and demonstrate that these amplitudes reproduce exactly what one expects from the dual symmetric orbifold perspective. }

\begin{document}
\maketitle
\flushbottom

\section{Introduction}

In recent years we have seen growing evidence for the equivalence between string theory on ${\rm AdS}_3\times {\rm S}^3 \times \mathbb{T}^4$ with one unit of NS-NS flux ($k=1$), and the symmetric orbifold theory of $\mathbb{T}^4$ \cite{Gaberdiel:2018rqv,Eberhardt:2018ouy,Eberhardt:2019ywk}, see also \cite{Giribet:2018ada}. 
In particular, it was shown in \cite{Eberhardt:2018ouy} that the spectra of the two descriptions agree precisely, and the structure of correlation functions has also been matched \cite{Eberhardt:2019ywk,Dei:2020zui}. 

The case with NS-NS flux $k=1$ is best established, but there is also good evidence that the long-string sector of the world-sheet theory with $k>1$ is dual to the symmetric orbifold of ${\cal N}=4$ Liouville theory times $\mathbb{T}^4$ \cite{Eberhardt:2019qcl}. Furthermore, this $k>1$ generalisation also seems to apply to the bosonic set-up for which there is a relation between string theory on ${\rm AdS}_3 \times X$ at the WZW point \cite{Maldacena:2000hw}, and the symmetric orbifold of Liouville theory times $X$ \cite{Eberhardt:2019qcl}. Some aspects of this bosonic correspondence were tested further in \cite{Dei:2019osr,Dei:2019iym}. In particular, it was shown in \cite{Dei:2019osr} that the null-states of the Liouville theory of the symmetric orbifold theory correspond to BRST exact states from the dual world-sheet perspective, and the general structure of twisted sector correlators (in particular, the differential equations that characterise them) was studied in \cite{Dei:2019iym}, see also \cite{Arutyunov:1997gt,Arutyunov:1997gi,Jevicki:1998bm,Lunin:2000yv,Lunin:2001pw,Pakman:2009zz,Roumpedakis:2018tdb} for earlier work on twisted sector correlators in symmetric orbifold theories. 

In this paper we subject the bosonic duality proposal to another consistency check. In reproducing the spectrum of the dual CFT from the world-sheet the identification of the scaling operator of the dual CFT, $L_0^{\rm CFT}$ with one of the $\mathfrak{sl}(2,\mathds{R})$ currents of the WZW model, 
$L_0^{\rm CFT}=J^3_0$, was used. This allowed one, for example, to calculate the full (single-particle) partition function of the symmetric orbifold theory from the world-sheet perspective \cite{Gaberdiel:2018rqv,Eberhardt:2018ouy}. However, we can also identify the symmetric orbifold stress-tensor with a specific world-sheet vertex operator, see \cite{Dei:2019osr}. Then we can calculate the $3$-point correlation functions of this vertex operator with, say, the vertex operators that correspond to the $w$-cycle twisted sector ground states of the dual CFT, from which we can also read off the conformal dimension of the latter. At least on the face of it, this gives a different way of determining these conformal dimensions, and it is the aim of this paper to confirm that this alternative method leads to the same result. Among other things, our analysis also confirms that the method of determining the correlators and the solution proposed in \cite{Eberhardt:2019ywk} is consistent with this constraint.\footnote{Previous work on calculating these spectrally flowed $\mathfrak{sl}(2,\mathds{R})$ correlation functions includes in particular \cite{Maldacena:2001km}, see also \cite{Ribault:2005ms,Giribet:2005mc,Baron:2008qf}.}
We also make a few technical advances in this paper. First of all, we clarify precisely which stress-energy operator in the dual CFT our world-sheet vertex operator corresponds to, see Section~\ref{sec:2.1} --- this is slightly subtle since the world-sheet theory only sees the single particle sector of the symmetric orbifold theory. Furthermore, in the process of determining the correlation functions we had to generalise the methods of \cite{Eberhardt:2019ywk} to allow also for descendant states, see Section~\ref{ward-identities-des}. While this is in principle straightforward, the analysis is actually quite complicated (and the problem becomes over-constrained), and it is very reassuring to see that this really works out as expected. 
\smallskip

The paper is organised as follows. We introduce our notation and set the stage for our calculation in Section~\ref{sec:2}. Section~\ref{sec:ward} reviews the derivation of the $\mathfrak{sl}(2,\mathds{R})$ Ward identities of \cite{Eberhardt:2019ywk}, and shows how they can be generalised to descendant states, see Section~\ref{ward-identities-des}. This is then applied to the correlators of interest in Section~\ref{sec:computations}. We conclude in Section~\ref{sec:concl}, and there are three appendices where some of the more technical material is described.

\section{The basic world-sheet set-up}\label{sec:2}

In this paper we shall consider bosonic string theory on $\text{AdS}_3 \times X$, for which the AdS factor can be described by an $\mathfrak{sl}(2,\mathds R)_k$ WZW model \cite{Maldacena:2000hw}. We shall work in the conventions in which the $\mathfrak{sl}(2,\mathds R)_k$ generators satisfy 
\begin{subequations}
\begin{align}
[J^3_m, J^3_n] &= -\tfrac{k}{2} \,  m \, \delta_{m+n,0} \ , \\
[J^3_m, J^\pm_n] &= \pm J^{\pm}_{m+n} \ , \\
[J^+_m, J^-_n] &= k \, m \, \delta_{m+n,0} - 2J^3_{m+n} \ .
\end{align}
\end{subequations}
Since the central charge of the $\mathfrak{sl}(2,\mathds R)_k$ algebra is $c=\frac{3k}{k-2}$, the  central charge of the internal CFT has to be 
\be 
c_X = 26-\frac{3k}{k-2} \ ,
\ee
so that the complete background is critical.

The states that reproduce the symmetric orbifold spectrum sit in the (spectrally flowed) \emph{continuous} representations of the $\mathfrak{sl}(2,\mathds R)_k$ algebra \cite{Gaberdiel:2018rqv,Giribet:2018ada}, and we shall denote the corresponding highest weight states by $|j,m\rangle$. Here $j=\frac{1}{2} + i s$ with $s\in\mathds{R}$ denotes the spin, while $m$ takes values in $m\in \mathbb{Z} + \lambda$, with $\lambda$ being an independent parameter of $j$ (or $s$). The Casimir of the $\mathfrak{sl}(2,\mathds R)$ representation depends on $j$ via 
\be 
C(j) = -j(j-1) \ ,
\ee
and we shall work in the conventions in which the modes of $\mathfrak{sl}(2,\mathds R)_k$ act on these states as
\begin{subequations}\label{hwrep}
\begin{align}
J^+_0 \ket{j,m} &= (m+j) \ket{j,m+1} \ , & J^+_n \ket{j,m} = 0 \ , \quad n> 0 \ , \\
J^3_0 \ket{j,m} &= m  \ket{j,m} \ , & J^3_n \ket{j,m} = 0 \ , \quad n> 0 \ , \\
J^-_0 \ket{j,m} &= (m-j) \ket{j,m-1} \ , & J^-_n \ket{j,m} = 0 \ , \quad n> 0 \ . 
\end{align} 
\end{subequations}
The dual CFT is not just the symmetric orbifold of $X$, but also involves a Liouville factor \cite{Eberhardt:2019qcl}. The stress-energy tensor of Liouville is strictly speaking not part of the Liouville spectrum, and this is reflected on the world-sheet in that it arises from a discrete representation with $j\in\mathds{R}$ and $j>\frac{1}{2}$ \cite{Dei:2019osr}, see eq.~(\ref{mj-choice}) below. 

The above description refers to the representations before spectral flow. The spectrally flowed representations are obtained from them by composing the action of the modes with the spectral flow automorphism
\be
\sigma^w (J^\pm_n) = J^\pm_{n\mp w} \ , \qquad  \sigma^w (J^3_n) = J^3_n + \tfrac{k\, w}{2}\delta_{n,0} \ .
\label{def-spectral-flow}
\ee
This is to say, on the states of the $w$-spectrally flowed representation $[\psi]^{(w)}$, the modes of the affine algebra act as 
\be 
J^a_n [\psi]^{(w)} \equiv [\sigma^w(J^a_n) \psi]^{(w)} \ ,
\label{action-spectral-flow}
\ee
where $\psi$ is a state in the highest weight representation eq.~(\ref{hwrep}), and $w\in\mathbb{N}$. 

\subsection{The symmetric orbifold stress-energy tensor}\label{sec:2.1}

We are interested in analysing correlation functions of the symmetric orbifold stress-energy tensor on the world-sheet. In order to be able to do so, we first need to identify the world-sheet state that corresponds to the spacetime stress-energy tensor. Actually, there are two natural stress-energy tensors that appear, namely the full stress-energy tensor of the symmetric orbifold theory, and the one that is just associated with the Liouville factor. Using the DDF operators of \cite{Giveon:1998ns,Eberhardt:2019qcl}, the relevant world-sheet states which we shall denote by $T=\mathcal{L}_{-2} |0\rangle$ and $T^\text{L}=\mathcal{L}^{\text{L}}_{-2} |0\rangle$, respectively, were worked out explicitly in \cite{Dei:2019osr}, and they take the form 
\begin{align}
T = & \Bigl[ c_{-} J_{-2}^- \ket{j,m+1} + c_{3} J^3_{-2} \ket{j,m}  + c_{+} J^+_{-2} \ket{j,m-1} \nonumber \\[-4pt] 
& \quad + c_{--} J_{-1}^-J_{-1}^- \ket{j,m+2} + c_{-3} J^-_{-1} J^3_{-1} \ket{j,m+1} \nonumber \\ 
& \quad + c_{-+} J^-_{-1}J^+_{-1} \ket{j,m} + c_{33} J^3_{-1}J^3_{-1} \ket{j,m} \nonumber \\[-6pt]
& \quad + c_{3+} J^3_{-1}J^+_{-1} \ket{j,m-1} + c_{++} J^+_{-1}J^+_{-1} \ket{j,m-2} \Bigr]^{(1)} \ , 
\label{stress-tensor-tot} \\[6pt]
T^\text{L} = & \Bigl[ c_{-}^{\textnormal{L}} J_{-2}^- \ket{j,m+1} +c_{3}^{\textnormal{L}} J^3_{-2} \ket{j,m} + c_{+}^{\textnormal{L}} J^+_{-2} \ket{j,m-1} \nonumber \\[-2pt]
& \quad + c_{--}^{\textnormal{L}} J_{-1}^-J_{-1}^- \ket{j,m+2} + c_{-3}^{\textnormal{L}} J^-_{-1} J^3_{-1} \ket{j,m+1} \nonumber \\ 
& \quad + c_{-+}^{\textnormal{L}} J^-_{-1}J^+_{-1} \ket{j,m} + c_{33}^{\textnormal{L}} J^3_{-1}J^3_{-1} \ket{j,m} \nonumber \\[-6pt] 
& \quad + c_{3+}^{\textnormal{L}} J^3_{-1}J^+_{-1} \ket{j,m-1} + c_{++}^{\textnormal{L}} J^+_{-1}J^+_{-1} \ket{j,m-2} - L_{-2}^X \ket{j,m}\Bigr]^{(1)} \ , 
\label{stress-tensor-L}
\end{align}
where 
\be 
m = 2-\frac{k}{2} \ , \qquad \qquad j = \frac{k-2}{2} \quad \text{or} \quad j = 1-\frac{k-2}{2} \ ,
\label{mj-choice}
\ee
and the coefficients $c_-, c_+, \dots$  and $c_-^{\text{L}}, c_+^{\text{L}}, \dots$ that appear  in eqs.~\eqref{stress-tensor-tot} and \eqref{stress-tensor-L}, respectively, are functions of $j,k$ and $m$, see Appendix \ref{app:stress-tensor-coefficients} for more details.  We should mention that for $k\neq 3$ --- this corresponds to the situation where the dual Liouville theory has a non-trivial background charge, see eq.~(2.42) of \cite{Eberhardt:2019qcl} --- the two choices $j = \frac{k-2}{2}$ vs.\ $j = 1-\frac{k-2}{2}$ are not exactly equivalent: while in general the $\mathfrak{sl}(2,\mathds{R})$ representation defined by eq.~(\ref{hwrep})  differs from the one where $j$ is replaced by $1-j$ only by an $m$-dependent rescaling of the basis vectors (and hence defines an equivalent representation), these rescaling factors become zero (or infinity) if $m\pm j=0$ or $m\pm (1-j)=0$, and this actually happens in the above case.\footnote{For example, for $j=\frac{k-2}{2}$, $m-1+j=0$, and thus this problem arises for the terms of the form $|j,m-1\rangle$ in eq.~(\ref{stress-tensor-tot}) and (\ref{stress-tensor-L}).} For most of the calculations we will perform below, the results will actually be independent of which choice we make, but for the case of the 2-point functions in Sections~\ref{sec:4.3.3} and \ref{sec:4.3.4} the two choices lead to slightly different normalisation factors. 

The above stress-energy tensors have central charges
\be\label{cseed}
c_{\text{seed}} = 6k \ , \qquad \qquad c^{\text L}_{\text{seed}} = c_{\text{seed}} - c_{X} = 1 + \frac{6(k-3)^2}{k-2} \ , 
\ee
where in either case the central charge refers to that of a single copy (i.e.\ the so-called seed theory) of the symmetric orbifold. These central charges appear in the $2$-point function of the corresponding stress-energy tensors, i.e.\ the dual CFT correlators have the form 
\be 
\left\langle T(x_1) T(x_2) \right\rangle = \frac{c_{\text{seed}}}{2(x_1 - x_2)^4}  \ , \qquad 
\left\langle T^{\text L}(x_1) T^{\text L}(x_2) \right\rangle = \frac{c^{\text L}_{\text{seed}}}{2(x_1 - x_2)^4}  \ .
\label{TT}
\ee
The attentive reader may be surprised that the central charge of the seed theory (rather than that of the full symmetric orbifold theory) appears here. The reason for this is that the world-sheet operator only sees the single-particle sector of the dual CFT, see also the discussion in Section~2.5.1 of \cite{Eberhardt:2019qcl}. More specifically, the $w$-twisted sector of the symmetric orbifold theory contains (for $w\geq 2$) two separate $\Delta h=2$ descendants, namely 
\be\label{des2}
\sum_{i=1}^{w} L^i_{-2} \sigma_w  \ , \qquad \hbox{and} \qquad \sum_{i=w+1}^{N} L^i_{-2} \sigma_w \ , 
\ee
where we have assumed that the corresponding permutation is simply $\pi_w=(12\cdots w)$. Only the first state is a single particle state --- and this is therefore the state that is dual to a world-sheet vertex operator --- while the actual stress-energy descendant of the twisted sector ground state is the sum of the two terms, see also the discussion around eq.~(5.13) in \cite{Gukov:2004ym}. The dual CFT state to the world-sheet state $T$ is the $w=1$ generalisation of the first state, and therefore describes in effect the stress-energy tensor of the seed theory.\footnote{For general $w\geq 2$, the centraliser of $\pi_w$ is $\mathbb{Z}_w \times S_{N-w}$, and thus both states in (\ref{des2}) are separately orbifold invariant. For $w=1$, on the other hand, the actual centraliser is $S_N$, and thus only the sum of the two states is really orbifold invariant. However, the sum of the two states has central charge $Nc_{\text{seed}}$, and therefore does not make sense in the large $N$ limit which is always implicit in the perturbative world-sheet description. The world-sheet calculation therefore picks out the analogue of the first term.} Obviously the same considerations apply to the correlators where we replace $T$ by $T^{\text{L}}$.

\medskip

The other family of correlators we shall be reproducing from the world-sheet are the dual CFT correlators of the form 
\be\label{sTs}
\left\langle \sigma_w(\infty) \, T(x) \,  \sigma_w(0) \right\rangle \ , \qquad \hbox{and} \qquad
\left\langle \sigma_w(\infty) \, T^{\text L}(x) \, \sigma_w(0) \right\rangle \ , 
\ee
where $\sigma_w$ denotes the ground state of the $w$-cycle twisted sector of the symmetric orbifold. We can calculate these correlators by going to the covering space, i.e.\ by rewriting the $x$ variables in terms of the covering map $z\mapsto \Gamma(z)$ with 
\be
 \Gamma(z) = x z^w \ . 
\ee
(This function maps $(0,1,\infty)$ to $(0,x,\infty)$, and has the correct branching behaviour near each of these points.) Then we can use the transformation property of the stress-energy tensor under general conformal tranformations, see e.g.\ \cite{Gaberdiel:1994fs}
\be\label{Ttrans}
T(x) = \frac{1}{\Gamma'(z)^2} \, \Bigl( T(z) - \frac{c_{\text{seed}}}{12} S[\Gamma(z)] \Bigr) \ , 
\qquad 
T^{\text{L}}(x) = \frac{1}{\Gamma'(z)^2} \, \Bigl( T^{\text{L}}(z) - \frac{c^{\text{L}}_{\text{seed}}}{12} S[\Gamma(z)] \Bigr) \ ,
\ee
where $S[\Gamma(z)]$ is the Schwarzian derivative, and the coefficient in front of $S[\Gamma(z)]$ comes from $L_2 T$ and $L_2 T^{\text{L}}$, respectively, where $L_n$ are the modes of $T$; this then leads to $c_{\text{seed}}$ and $c^{\text{L}}_{\text{seed}}$, respectively. Note that in either case the central charge that appears here is that of the seed theory, see the discussion around eq.~(\ref{TT}). 

For the above choice of $\Gamma(z)$ the Schwarzian derivative equals 
\be
S[\Gamma(z)] = \frac{\Gamma'''(z)}{\Gamma'(z)} - \frac{3}{2} \Biggl( \frac{\Gamma''(z)}{\Gamma'(z)} \Biggr)^2 = - \frac{(w^2-1)}{2z^2} \ . 
\ee
After we have applied the covering map, the ground state $\sigma_w$ disappears, and thus only the second term in (\ref{Ttrans}) contributes. Thus the correlator equals (upon setting $z=1$)
\be\label{sTsans}
\left\langle \sigma_w(\infty) \, T(x) \,  \sigma_w(0) \right\rangle = \frac{c_{\text{seed}} (w^2-1)}{24\, w^2}  \, \frac{1}{x^2} \ . 
\ee
Similarly, we find for the correlator where we have replaced $T$ by $T^{\text L}$, eq.~(\ref{sTsans}) with $c_{\text{seed}}^{\text L}$ in place of $c_{\text{seed}}$. 
\medskip

It is the aim of this paper to calculate the correlators (\ref{TT}) and (\ref{sTs}) from the world-sheet perspective. At least on the face of it, this is a fairly non-trivial consistency check on the calculation of the world-sheet correlators (and the specific solution proposed in \cite{Eberhardt:2019ywk}, see eq.~(\ref{solution-equation-5.8})), as well as more generally of the precise duality proposal.  

Note that for the world-sheet calculation of the  $3$-point functions in (\ref{sTs}) we also need to identify the world-sheet state that corresponds to the $w$-cycle twisted sector ground state; this was already done in \cite{Eberhardt:2019ywk}, and it is a $w$-spectrally flowed affine highest weight state, i.e.\ a state of the form $[|j,m\rangle]^{(w)}$, where $j=\frac{k-2}{2}$ or $j=1 - \frac{k-2}{2}$, and $m$ is determined by the mass-shell condition.

\subsection[The $x$-dependence of the vertex operators]{The $\boldsymbol{x}$-dependence of the vertex operators}

One important subtlety that appears in the calculation of these correlation functions is that the vertex operators on the world-sheet naturally depend on two kinds of coordinates \cite{Eberhardt:2019ywk}. First of all, the vertex operators depend on the usual position $z$ where they are inserted on the world-sheet. However, it is also natural to make them depend on a coordinate $x$ that refers to the position on the boundary sphere where the corresponding dual CFT state is inserted. In fact, there is a canonical way in which these dependencies can be determined: starting from the identification of vertex operators and states at, say, $z=x=0$, the dependence on the two coordinates is determined by conjugation with the corresponding translation operators, 
\be
V(\psi;x,z) = e^{z L_{-1}} \, e^{x J^+_0} \, V(\psi;0,0) \, e^{-x J^+_0} \, e^{-z L_{-1}} \ , 
\ee
where $L_{-1}$ is the usual translation operator on the world-sheet, and we have used the  identification $J^+_0 \cong L_{-1}^{\text{CFT}}$. While this may seem overly pedantic, it actually has a significant impact on the definition of the vertex operators associated to spectrally flowed states. The basic reason for this is that in defining the spectral flow automorphism, see eq.~(\ref{def-spectral-flow}), a preferential role is given to the $J^3_0$ generator of $\mathfrak{sl}(2,\mathds{R})$, but this choice is not invariant under conjugation with respect to $e^{x J^+_0}$, and as a consequence the `direction' of spectral flow depends on $x$ in a non-trivial manner.\footnote{Most previous work on spectrally flowed $\mathfrak{sl}(2,\mathds{R})$ correlators, see e.g.\ \cite{Ribault:2005ms,Giribet:2005mc,Baron:2008qf}, has set $x=0$ uniformly, which is not appropriate in our context.}  In fact, this can be seen very explicitly by considering the OPE of the $\mathfrak{sl}(2,\mathds{R})$ currents with spectrally flowed vertex operators. Writing 
\be
V^w_h(x;z) \equiv V\bigl( [ |j,m\rangle]^{(w)};x,z \bigr) \ , \qquad \hbox{with} \quad h = m +  \tfrac{kw}{2} \ , 
\ee
and suppressing the $j$-dependence, we have the OPE relations 
\begin{subequations}
\begin{align}
J^+(z) V_{h}^{w}(x;\zeta) & \sim  (h-\tfrac{k w}{2} + j) \frac{V_{h+1}^{w}(x;\zeta)}{(z-\zeta)^{w+1}} 
+ \sum_{l=1}^{w-1} \frac{J^+_l V_h^w(x;\zeta)}{(z-\zeta)^{l+1}} 
+ \frac{\partial_x V^{w}_{h}(x;\zeta)}{(z-\zeta)} \ , \\
J^3(z) V_{h}^{w}(x;\zeta) & \sim 
x \, (h-\tfrac{k w}{2} + j) \frac{V_{h+1}^{w}(x;\zeta) }{(z-\zeta)^{w+1}} 
+ x \, \sum_{l=1}^{w-1} \frac{J^+_l V_h^w(x;\zeta)}{(z-\zeta)^{l+1}} 
+ \frac{(h +x \partial_x)V^{w}_{h}(x;\zeta) }{(z-\zeta)} \ , \\
J^-(z) V_{h}^{w}(x;\zeta) & \sim  x^2 \, (h-\tfrac{k w}{2} + j) \frac{V_{h+1}^{w}(x;\zeta)}{(z-\zeta)^{w+1}} + x^2 \,  \sum_{l=1}^{w-1} \frac{J^+_l V_h^w(x;\zeta)}{(z-\zeta)^{l+1}} 
 \nonumber \\ & \quad  
+  \frac{(2hx + x^2 \partial_x) V^{w}_{h}(x;\zeta)}{(z-\zeta)} \ , \label{eq:OPE-Jm-Vxz1}
\end{align}
\label{eq:OPE-Jm-Vxz}
\end{subequations}

\noindent where we have only written out the singular terms. They follow from the `usual' OPE relations\footnote{In order to obtain these formulae we expand the currents in terms of modes as $J(z) = \sum_{n} J_n z^{-n-1}$, and then use the action defined by (\ref{action-spectral-flow}).\label{foot1}} at $z=x=0$, 
\begin{subequations}
\begin{align}
J^+(z)V^{w}_{h}(0;0) & \sim (h-\tfrac{kw}{2}+j) \frac{V_{h+1}^w(0;0)}{z^{w+1}} + \sum_{l=1}^{w-1} \frac{J^+_l V_h^w(0;0)}{z^{l+1}} +  \frac{\partial_x V_h^w(0;0)}{z} \ , \\
J^3(z)V^{w}_{h}(0;0) & \sim \frac{h \, V_h^w(0;0)}{z} \ , \\
J^-(z)V^{w}_{h}(0;0) & \sim (h-\tfrac{kw}{2}-j) \, V_{h-1}^w(0;0) \, z^{w-1} + \mathcal{O}(z^w)
\label{eq:OPE-Jm-V00}
\end{align}
\end{subequations}
by using the conjugation action of $e^{x J^+_0}$ on the currents, see \cite{Eberhardt:2019ywk} for more details.

\section{The Ward identities on the world-sheet}
\label{sec:ward}

The basic strategy to determine the world-sheet correlators is to use the Ward identities associated to the $\mathfrak{sl}(2,\mathds{R})_k$ currents. This method was developed for correlators involving only primary states in \cite{Eberhardt:2019ywk}, see also \cite{Dei:2020zui} for a generalisation to the free field realisation of $\mathfrak{psu}(1,1|2)_1$; this general method will be briefly reviewed  in Section~\ref{Ward-identities-hws}. For the application we have in mind we also need to determine the Ward identities for correlators involving descendant states, and we explain in Section~\ref{ward-identities-des} the modifications that are required for that case.

\subsection{Ward identities for correlators of highest weight states}
\label{Ward-identities-hws}

Let us start with reviewing how the Ward identities for the correlators of the form 
\be
\left \langle \prod_{i=1}^{n} V^{w_i}_{h_i} \left(x_i;z_i\right) \right\rangle 
\label{naked-correlator}
\ee
can be derived. Given the OPEs of the $\mathfrak{sl}(2,\mathds{R})$ currents with these vertex operators, see eq.~(\ref{eq:OPE-Jm-Vxz}), we can determine the correlators where we insert a current $J^a(z)$ into the above correlator. Since all the singular terms are either explicitly known, or involve the action of a $J^+_\ell$ mode, we can thereby express all of these correlators in terms of the `unknown' correlators 
\be
F^i_\ell \equiv \left\langle [J^+_{\ell}V^{w_i}_{h_i}] \left(x_i;z_i\right)\, \prod_{j\neq i} V^{w_j}_{h_j} \left(x_j;z_j\right) \right\rangle \ ,
\label{eq:Fil}
\ee 
where $\ell=1,\ldots,w_i-1$, and $i=1,\ldots,n$. The key idea by means of which one can determine these unknowns is to use the fact that the OPE \eqref{eq:OPE-Jm-V00} is regular, i.e.\ that the singular terms in (\ref{eq:OPE-Jm-Vxz1}) simply arise because of the conjugation with $e^{x J^+_0}$. This allows us to remove these singular terms for any fixed $j$; for example, for $z\sim z_j$ the combination
\begin{multline}
\left\langle \Bigl( J^-(z) - 2x_j J^3(z) + x_j^2 J^+(z) \Bigr) \prod_{i=1}^n V_{h_i}^{w_i}(x_i;z_i) \right\rangle \\
= (h_j - \tfrac{k w_j}{2} + j_j) \left\langle V_{h_j-1}^{w_j}(x_j;z_j) \prod_{i \neq j} V_{h_i}^{w_i}(x_i;z_i)\right\rangle (z-z_j)^{w_j-1} + \mathcal{O}\bigl( (z-z_j)^{w_j} \bigr) \ ,
\label{eq:to-compare}
\end{multline}
is regular. On the other hand, as mentioned before we can also compute the left-hand-side directly using the 
OPEs \eqref{eq:OPE-Jm-Vxz}, and we find 
\begin{multline}
\left\langle \Bigl( J^-(z) - 2x_j J^3(z) + x_j^2 J^+(z) \Bigr) \prod_{i=1}^n V_{h_i}^{w_i}(x_i;z_i) \right\rangle \\
= \sum_{i \neq j} \Biggl( \frac{2(x_i-x_j)h_i + (x_i-x_j)^2 \partial_{x_i}}{(z-z_i)} \Bigl\langle \prod_{l=1}^n V^{w_l}_{h_l}(x_l;z_l) \Bigr\rangle + \sum_{\ell=1}^{w_i-1} \frac{(x_i-x_j)^2}{(z-z_i)^{\ell + 1}} F^i_\ell  \\
+ (h_i -\tfrac{k w_i}{2} + j_i)\frac{(x_i-x_j)^2}{(z-z_i)^{\ell + 1}} \Bigl\langle V^{w_i}_{h_i+1}(x_i;z_i) \prod_{l\neq i} V^{w_l}_{h_l}(x_l;z_l) \Bigr\rangle \Biggr) \ , 
\label{eq:to-expand}
\end{multline}
where $F^i_\ell$ has been defined in \eqref{eq:Fil}. Requiring that this expression is of the form of eq.~(\ref{eq:to-compare}), i.e.\  Taylor expanding \eqref{eq:to-expand} for $z \to z_j$ and comparing it with \eqref{eq:to-compare}, then leads to $w_j$ identities for each $j$. In particular, the first $w_j-1$ coefficients define a homogeneous linear system for the unknowns $F^i_\ell$ with $\ell \in \{1, \dots, w_i-1 \}$; this linear system has as many equations as unknowns, namely $\sum_{i=1}^n (w_i-1)$, and a non-trivial solution exists provided that \cite{Eberhardt:2019ywk}
\be 
\sum_{i \neq j} w_i \geq w_j -1 
\ee 
for all $j$. In this case, all the $F^i_\ell$ can be written in terms of the correlators in \eqref{naked-correlator} (and vice versa). Moreover, by comparing the terms of order $(z-z_j)^{w_j-1}$ in the Taylor expansion of \eqref{eq:to-compare} and \eqref{eq:to-expand}, recursion relations for correlators with shifted values of the $h_i$ can be derived, see  \cite{Eberhardt:2019ywk} for more details. 
In general,  these recursion relations are complicated to solve, but provided that\footnote{This is the condition for the case of a $3$-point function. There are also some constraints on the spectral flow labels $w_i$, see eq.~(5.2) in \cite{Eberhardt:2019ywk}, but they will always be satisfied for us.} 
\be 
j_1 + j_2 + j_3 = \frac{k}{2} \ , 
\label{j-constraint}
\ee  
a simple solution exists \cite{Eberhardt:2019ywk} 
\begin{multline}
\langle V^{w_1}_{h_1}\left(x_1;z_1\right) V^{w_2}_{h_2}\left(x_2;z_2\right)V^{w_3}_{h_3}\left(x_3;z_3\right) \rangle \\
= C(j_1,j_2,j_3) \prod_{i=1}^{3} (a_i^{\Gamma})^{-h_i}\prod_{i\neq j} (z_i-z_j)^{\Delta_l^0 - \Delta_i^0 - \Delta_j^0} \ ,
\label{solution-equation-5.8}
\end{multline}
where the coefficients $a_i^\Gamma$ (that are determined by the corresponding covering map) read explicitly \cite{Lunin:2000yv,Pakman:2009zz,Eberhardt:2019ywk}
\begin{equation}
a_i^{\Gamma} = \frac{\left( \begin{array}{c} \tfrac{1}{2}(w_{i}+w_{i+1}+w_{i+2}-1) \\
\tfrac{1}{2}(-w_{i}+w_{i+1}+w_{i+2}-1) \end{array} \right)}{\Biggl( \begin{array}{c} \tfrac{1}{2}(-w_{i}+w_{i+1}-w_{i+2}-1) \\ \tfrac{1}{2}(w_{i}+w_{i+1}-w_{i+2}-1) \end{array} \Biggr)}\, \frac{(x_{i}-x_{i+1})(x_{i+2}-x_{i})(z_{i+1}-z_{i+2})^{w_i}}{(x_{i+1}-x_{i+2})(z_{i}-z_{i+1})^{w_i}(z_{i+2}-z_{i})^{w_i}} \ ,
\label{a-G-def}
\end{equation}
and the indices in eq.~(\ref{a-G-def}) are to be understood mod 3. Note that the last factor in (\ref{solution-equation-5.8}) just reproduces the usual $z_i$ dependence of a $3$-point function of quasi-primary fields, where the relevant (world-sheet) conformal dimensions are $\Delta^0_i = \Delta_i + w_i \, h_i$, and $\Delta_l^0$ is the conformal dimension associated to the `third' field, i.e.\ the one which is neither $i$ nor $j$. 

\subsection{Ward identities for correlators of descendants}
\label{ward-identities-des}

For the analysis of the correlation functions we are interested in, we also need to compute correlation functions of the form (\ref{eq:Fil}) where yet an additional current $J^a(z)$ has been inserted. These correlators can be determined by considering the OPEs 
\begin{subequations}
\begin{align}
J^+(z) \, [J^+_lV^{w_i}_{h_i}](x_i;z_i) &\sim \sum_{m=1}^{w_i} \frac{[J^+_lJ^+_mV^{w_i}_{h_i}](x_i;z_i)}{(z-z_i)^{m+1}} + \frac{\partial_{x_i}   \, [J^+_lV^{w_i}_{h_i}](x_i;z_i) }{(z-z_i)}   \ , \\
J^3(z) \, [J^+_lV^{w_i}_{h_i}](x_i;z_i) & \sim 
 \sum_{m=1}^{w_i-l} \frac{[J_{m+l}^+V^{w_i}_{h_i}](x_i;z_i)}{(z-z_i)^{m+1}} +x_i \sum_{m=1}^{w_i} \frac{[J^+_lJ^+_mV^{w_i}_{h_i}](x_i;z_i)}{(z-z_i)^{m+1}}  \nonumber \\
& \qquad + \frac{\left((h_i+1)+x_i\partial_{x_i}\right) \, [J^+_l V^{w_i}_{h_i}](x_i;z_i)}{(z-z_i)}  \ , \\[4pt] 
J^-(z) \, [J^+_l V^{w_i}_{h_i}](x_i;z_i) & \sim  
 2x_i\sum_{m=1}^{w_i-l} \frac{[J_{m+l}^+V^{w_i}_{h_i}](x_i;z_i)}{(z-z_i)^{m+1}} + x_i^2 \sum_{m=1}^{w_i} \frac{[J^+_lJ^+_mV^{w_i}_{h_i}](x_i;z_i)}{(z-z_i)^{m+1}} \nonumber \\
& \qquad + \frac{\left(2x_i(h_i+1)+x_i^2\partial_{x_i}\right) \, [J^+_lV^{w_i}_{h_i}](x_i;z_i) }{(z-z_i)} \ ,
\end{align}
\label{OPEs-with-descendants}
\end{subequations}

\noindent that can be derived as before, see footnote~\ref{foot1}, but now taking into account that the vertex operator $[J^+_lV^{w_i}_{h_i}](x_i;z_i)$ is associated with a descendant state. However, as is clear from the right-hand-sides of the above expressions, more complicated  `unknowns' will appear in the process, in particular, 
\begin{align}
G_i\left(\ell,m\right) & \equiv \Bigl\langle [J^+_\ell J^+_m V^{w_i}_{h_i}](x_i;z_i)\, \prod_{r\neq i}^{n} V_{h_r}^{w_r}\left(x_r;z_r\right) \Bigr\rangle \ ,  \label{Gilm}\\
M_{i,p}\left(\ell,m\right) & \equiv \Bigl\langle [J^+_\ell V^{w_i}_{h_i}](x_i;z_i) \, [J^+_m V^{w_p}_{h_p}](x_p;z_p) \, \prod_{r\neq i,p}^{n}V^{w_r}_{h_r}(x_r;z_r) \Bigr\rangle \ . \label{Miplm}
\end{align}
In order to determine them, we consider the correlator
\be
\Bigl\langle \left(J^-(z)-2x_jJ^3(z)+x_j^2J^+(z) \right) \, [J^+_\ell V^{w_i}_{h_i}](x_i;z_i)\, \prod_{r\neq i}^{n} V^{w_r}_{h_r}\left(x_r;z_r\right) \Bigr\rangle \ ,
\label{correlator-combination-descendant}
\ee
where $\ell \in \{1,\ldots,w_i-1\}$, and analyse it in two different ways. First we evaluate it directly, using (\ref{OPEs-with-descendants}), and thereby obtain 
\begin{align}
& \Bigl\langle \left(J^-(z)-2x_jJ^3(z)+x_j^2J^+(z) \right) \, [J^+_\ell V^{w_i}_{h_i}] \left(x_i;z_i\right)\, \prod_{r\neq i}^{n} V^{w_r}_{h_r}\left(x_r;z_r\right) \Bigr\rangle \\
& =\frac{\left( 2(h_i+1)(x_i-x_j)+(x_i-x_j)^2\partial_{x_i} \right) \, F^i_\ell}{(z-z_i)} \nonumber \\ 
& \qquad + \sum_{m=1}^{ w_i-\ell } \frac{2(x_i-x_j) \, F^i_{\ell+m}}{(z-z_i)^{m+1}} +  \sum_{m=1}^{w_i} \frac{(x_i-x_j)^2 \, G_i(\ell,m)}{(z-z_i)^{m+1}} \nonumber \\ 
& \qquad +\sum_{p\neq j,i}\left(\frac{ \left( 2(x_p-x_j)h_p+(x_p-x_j)^2 \partial_{x_p} \right) F^i_\ell}{(z-z_p)}  + \sum_{m=1}^{w_p}\frac{(x_p-x_j)^2 \, M_{i,p}(\ell,m)}{(z-z_p)^{m+1}}  \right) \ .
\label{constraint-type-2-OPEs}
\end{align}
On the other hand, we know that as $z \to z_i$ the correlator must behave as 
\begin{multline}
\Bigl\langle \left(J^-(z)-2x_iJ^3(z)+x_i^2J^+(z) \right) \, [J^+_\ell V^{w_i}_{h_i}](x_i;z_i)\, \prod_{r \neq i}^{n} V^{w_r}_{h_r}\left(x_r;z_r\right) \Bigr\rangle \\ 
= \left(2h_i-k \ell \right)\Bigl\langle V^{w_i}_{h_i}\left(x_i;z_i\right) \prod_{r\neq i}^{n} V^{w_r}_{h_r}\left(x_r;z_r\right)\Bigr\rangle \, (z-z_i)^{\ell-1} + \mathcal{O} \left((z-z_i)^{\ell}\right) \ ,
\label{RRji}
\end{multline}
while for $z \to z_p$ with $p \neq i$ we find instead
\begin{align} 
& \Bigl\langle \left(J^-(z)-2x_pJ^3(z)+x_p^2J^+(z) \right) \, [J^+_\ell V^{w_i}_{h_i}](x_i;z_i) \, \prod_{r \neq i} V^{w_r}_{h_r}\left(x_r;z_r\right) \Bigr\rangle  \nonumber \\ 
& \qquad = \left(h_p-\tfrac{k w_p}{2}-j_p\right) \, 
\Bigl\langle [J^+_\ell V^{w_i}_{h_i}](x_i;z_i) \, V^{w_p}_{h_p-1}\left(x_p;z_p\right)\, \prod_{r \neq i,p}^{n} V^{w_r}_{h_r}\left(x_r;z_r\right) \Bigr\rangle (z-z_p)^{w_p-1} \nonumber \\ 
& \qquad \quad 
+ \mathcal{O}\left((z-z_p)^{w_p}\right) \ . \label{RRjp}
\end{align}
 As before, we can thus compare the Taylor expansion of \eqref{constraint-type-2-OPEs} for $z \to z_j$ for $j=i$ and $j \neq i$ with eqs.~\eqref{RRji} and \eqref{RRjp}, respectively. For each choice of $\ell\in\{1,\ldots,w_i-1\}$, we get $\ell$ relations of the first kind, and $w_j+1$ of the second,  and they relate $G_i(\ell,m)$ and $M_{i,p}(\ell,m)$ to $F_\ell^i$ and correlators without any insertions of $J^+_\ell$ modes.\footnote{In both cases, some of the $h_j$ values may be shifted.} Since the latter have already been determined, this allows us then to solve for all $G_i(\ell,m)$ and $M_{i,p}(\ell,m)$; in fact, the problem is overdetermined, and it is a non-trivial consistency condition that a solution exists at all.

\section{Explicit computations and results}
\label{sec:computations}

With these preparations at hand, we can now evaluate the correlation functions that are of interest to us: the world-sheet correlator that corresponds to (\ref{TT}) is 
\be 
\left\langle V \left(T;x_1,z_1 \right) V \left(T;x_2,z_2 \right) \right\rangle  
\ , 
\label{world-sheet-TT}
\ee
where $j_1$ and $j_2$ are chosen as in \eqref{mj-choice}; and the world-sheet correlator that corresponds to (\ref{sTs}) is 
\be
\langle V^{w}_{h_1}\left(0;0\right) \, V \left(T;x,z \right) \, V^{w}_{h_3}\left(\infty;\infty\right) \rangle
\ ,
\label{ratio-stress-tensor}
\ee
where 
\be 
h_1 = h_3 = \frac{6k \, (w^2-1)}{24 w} \ .
\ee
Since the normalisation of the correlators is not fixed by the Ward identities, we must in each case divide by the corresponding vacuum correlators, where we replace $V \left(T;x,z \right)$ by 
$V^{1}_{0}\left(x;z \right)$, i.e.\ the vertex operator that corresponds to the vacuum in the dual CFT. Note that the vacuum vertex operator arises for the same values of $m$ and $j$ as the stress-energy tensors, see eq.~(\ref{mj-choice}); as a consequence the $j$-dependent normalisation factors of the correlators, see e.g.\ eq.~(\ref{solution-equation-5.8}), drop out of this ratio. 

In the above formulae, we have written these correlators for the case of the full stress-energy tensor of the dual symmetric orbifold, but we will also be considering the corresponding correlators where $T$ is replaced by $T^{\text{L}}$, the stress-energy tensor associated to the Liouville factor. It is the aim of this section to explain in some detail how these world-sheet calculations can be performed.

\subsection{Decoupling spurious states}

In order to simplify the calculation, we first recall that the world-sheet theory contains spurious states that decouple from all correlation functions. They take the form \cite{Green:1987sp,Dei:2019osr}
\begin{align}
\ket{\psi} = L_{-1} \ket{\chi_1} +  \left( L_{-2} + \frac{3}{2} L^2_{-1}\right)\ket{\chi_2} \ ,
\label{spurious}
\end{align}
where
\begin{align}
\ket{\chi_1} &= 4 a \, (j-m-1) \, J_{-1}^{3}\ket{j,m}+4b \, (j-m-1) J_{-1}^+\ket{j,m-1} 
+ a \,  (k+2 m) J_{-1}^-\ket{j,m+1}  , \nonumber \\
\ket{\chi_2} &= \ket{j,m} \ ,
\label{spurious-chi}
\end{align}
and $a$ and $b$ are arbitrary parameters.\footnote{For some of these states we have also checked explicitly that they vanish in correlation functions, using the techniques of this paper. A more general argument for this was already given in \cite{Dei:2019osr}.} As a consequence, we are free to add suitable multiples of these spurious states to $T$ and $T^{\text{L}}$ in order to simplify the above expressions for them. In particular, we can replace in this manner the contribution of the $J^-_{-1}J^3_{-1}$, $J^-_{-1}J^+_{-1}$ and the $J^-_{-1}J^-_{-1}$ terms in $T$ and $T^{\text{L}}$ at the cost of modifying the coefficients of the other terms entering in \eqref{stress-tensor-tot} and \eqref{stress-tensor-L}. This allows us to bring $T$ and $T^{\text{L}}$ into the form 
\begin{align}
T = & \, c_1 \, J^3_{-1}J^+_{-1} \ket{j,1-\tfrac{k}{2}} + c_2 \, J^3_{-1}J^3_{-1} \ket{j,2-\tfrac{k}{2}} +c_3 \, J^3_{-2}\ket{j,2-\tfrac{k}{2}}\nonumber \\
& +c_4 \, J^-_{-1}J^+_{-1} \ket{j,2-\tfrac{k}{2}} +c_5 \, J^+_{-2}\ket{j,1-\tfrac{k}{2}} + c_6 \, L_{-2}^X\ket{j,2-\tfrac{k}{2}} \ ,
\label{stress-tensor-tot-simplified} \\
T^{\text{L}} = & \, c_1^{\textnormal{L}} \, J^3_{-1}J^+_{-1} \ket{j,1-\tfrac{k}{2}} + c_2^{\textnormal{L}} \, J^3_{-1}J^3_{-1} \ket{j,2-\tfrac{k}{2}} +c_3^{\textnormal{L}} \, J^3_{-2}\ket{j,2-\tfrac{k}{2}}\nonumber \\
& +c_4^{\textnormal{L}} \, J^-_{-1}J^+_{-1} \ket{j,2-\tfrac{k}{2}}  +c_5^{\textnormal{L}} \, J^+_{-2}\ket{j,1-\tfrac{k}{2}} + c_6^{\textnormal{L}} \, L_{-2}^X\ket{j,2-\tfrac{k}{2}} \ ,
\label{stress-tensor-L-simplified}
\end{align}
where $j$ is either  $j = \tfrac{k-2}{2}$ or $j = 1 - \tfrac{k-2}{2}$. The explicit form of the coefficients $c_1, \dots$ and $c^{\textnormal{L}}_1, \dots$ are spelled out in Appendix \ref{app:stress-tensor-coefficients}.  

\subsection{Wrapping of modes}
\label{sec:wrapping-of-modes}

We begin by discussing the correlators of the form \eqref{ratio-stress-tensor}, which we can write as a sum of terms of the form 
\be
\langle V^{w}_{h_1}\left(x_1;z_1 \right) V \bigl(\left[\ket{\phi} \right]^{(1)};x_2,z_2 \bigr) V^{w}_{h_3}\left(x_3;z_3\right)\rangle \ ,
\label{3pt-function-general}
\ee
where the state $\ket{\phi}$ stands for any of the terms appearing in eqs.~\eqref{stress-tensor-tot-simplified} or \eqref{stress-tensor-L-simplified}.  For concreteness, let us discuss the case $\ket{\phi} = J^3_{-2}\ket{j,m}$ with $j=\frac{k-2}{2}$ and $m=2-\frac{k}{2}$. Since the spectral flow of $J^3_{-2}$ is trivial, we have $\left[J^3_{-2}\ket{j,m} \right]^{(1)} = J^3_{-2}\left[\ket{j,m} \right]^{(1)}$, and we can write the $J^3_{-2}$ mode in terms of a contour integral, which we can then wrap around the other insertion points  
\begin{align}
& \langle V^{w}_{h_1}\left(x_1;z_1 \right) V \bigl( \left[J^3_{-2} \ket{j,m} \right]^{(1)} ; x_2,z_2\bigr) V^{w}_{h_3}\left(x_3;z_3\right)\rangle \\ 
&\qquad = \oint_{z_2} \frac{dz}{(z-z_2)^2} \langle \left(J^{3}(z)-x_2J^+(z) \right) \, V^{w}_{h_1}\left(x_1;z_1 \right) V \bigl( \left[ \ket{j,m} \right]^{(1)};x_2,z_2 \bigr) 
V^{w}_{h_3}\left(x_3;z_3\right)\rangle \nonumber \\ 
&\qquad  = -\sum_{i=1,3}\oint_{z_i} \frac{dz}{(z-z_2)^2}\langle \left(J^{3}(z)-x_2J^+(z) \right)V^{w}_{h_1}\left(x_1;z_1 \right) V_{h_2}^1 \left(x_2;z_2\right) V^{w}_{h_3}\left(x_3;z_3\right) \rangle \ , \nonumber
\label{J3-2-intermediate}
\end{align}
where in the final step we have rewritten $V \bigl( \left[ \ket{j,m} \right]^{(1)};x_2,z_2 \bigr)=V_{h_2}^1 \left(x_2;z_2\right)$ since this is now the $w=1$ spectrally flowed image of a highest weight state (with $j=\frac{k-2}{2}$). The OPEs of $J^3(z)$ and $J^+(z)$ near $z=z_1$ and $z=z_3$ can now be determined from eq.~\eqref{eq:OPE-Jm-Vxz}, and this leads to 
\begin{align}
& \langle V^{w}_{h_1}\left(x_1;z_1 \right) V \bigl(\left[J^3_{-2}\ket{j,m} \right]^{(1)} ;x_2,z_2\bigr) V^{w}_{h_3}\left(x_3;z_3\right)\rangle \nonumber \\[4pt]
& \qquad  = -\sum_{i=1,3} \Bigl[\tfrac{h_i+(x_i-x_2)\partial_{x_i}}{(z_i-z_2)^2} \langle V^{w}_{h_1}\left(x_1;z_1 \right) V^1_{h_2} \left(x_2;z_2\right) V^{w}_{h_3}\left(x_3;z_3\right)\rangle   \\ 
& \hspace{60pt} +  (x_i-x_2)\sum_{\ell=1}^{w}\frac{(-1)^\ell (\ell+1)}{(z_i-z_2)^{\ell+2}} \langle \left(J^+_\ell V^{w}_{h_i}\right) \left(x_i;z_i\right)  V^1_{h_2} \left(x_2;z_2\right) V^{w}_{h_p}\left(x_p;z_p\right)\rangle \Bigr] \ , \nonumber
\label{J3-2-final}
\end{align}
where $p\in \{1,3\}$ with $p\neq i$. The other terms can be computed in a similar manner, albeit that analysis is more complicated since for some of the terms that involve two $J^a$ modes, we will also need the OPEs of the form (\ref{OPEs-with-descendants}), and thus also the more complicated unknowns (\ref{Gilm}) and (\ref{Miplm}) will appear. We have collected the explicit results of these calculations in Appendix~\ref{app:formulas-for-additional-modes}.

\subsubsection[The $2$-point case]{The $\boldsymbol{2}$-point case}

The analysis of the correlators of the form (\ref{world-sheet-TT}) is similar, except that now we have a sum of contributions of the form 
\be
\langle V \bigl(\left[\ket{\phi}_1 \right]^{(1)};x_1,z_1 \bigr) \, V \bigl(\left[\ket{\phi}_2 \right]^{(1)};x_2,z_2 \bigr)  \rangle \ , 
\label{2pt-function-general}
\ee
where $\ket{\phi}_1$ and $\ket{\phi}_2$ stand for any of the terms appearing in eqs.~\eqref{stress-tensor-tot-simplified} or \eqref{stress-tensor-L-simplified}. For example, let us discuss the case with $\ket{\phi}_1 = J^+_{-2} \ket{j_1,m}$ and $\ket{\phi}_2= J^+_{-2} \ket{j_2,m}$, where our choice for $j_1$ and $j_2$ will be discussed below, see eq.~(\ref{jchoice}). Since $\sigma(J^+_{-1}) = J^+_{-2}$, we have $\left[J^+_{-2}\ket{j,m-1}\right]^{(1)}= J^+_{-1} \, \left[\ket{j,m-1}\right]^{(1)}$, and the correlator becomes\footnote{We work with the convention that $m=2-\frac{k}{2}$; then the $m$ eigenvalue of the term involving $J^+_{-2}$ in \eqref{stress-tensor-tot-simplified} or \eqref{stress-tensor-L-simplified} is $m-1$.}
\begin{align}
& \Bigl\langle V\bigl( \left[J^+_{-2} \ket{j_1,m-1}\right]^{(1)};x_1,z_1 \bigr) \, 
V\bigl(\left[J^+_{-2} \ket{j_2,m-1}\right]^{(1)};x_2,z_2 \bigr) \Bigr\rangle  \\
&\qquad = \oint_{z_1} \frac{dz}{(z-z_1)} \Bigl\langle J^+(z) V_{h-1}^1\left(x_1;z_1\right)V\bigl( J^+_{-1}\left[\ket{j_2,m-1}\right]^{(1)};x_2,z_2 \bigr) \Bigr\rangle \nonumber \\
& \qquad =  \frac{(h-1-\frac{k}{2}+j_2)}{(z_2-z_1)^2}  \Bigl\langle V_{h-1}^1\left(x_1;z_1\right) V\bigl(J^+_{-1}\left[\ket{j_2,m}\right]^{(1)};x_2,z_2 \bigr) \Bigr\rangle  \nonumber \\ 
& \qquad \qquad -  \frac{1}{(z_2-z_1)} \Bigl\langle V_{h-1}^1\left(x_1;z_1\right)\, 
V\bigl(J^+_{-1}J^+_{0}\left[\ket{j_2,m-1}\right]^{(1)}; x_2,z_2 \bigr) \Bigr\rangle  \ ,
\end{align}
where we have first written the $J^+_{-1}$ mode in terms of a contour integral around $z_1$, and then evaluated the contour integral by considering the residue at $z=z_2$. In a second step we then repeat the same procedure for the $J^+_{-1}$ mode acting on the vertex operator at $z=z_2$. In the end this leads to 
\begin{align}
&\Bigl\langle V\bigl(\left[J^+_{-2}\ket{j_1,m-1}\right]^{(1)}; x_1,z_1 \bigr)\, 
V\bigl(\left[ J^+_{-2} \ket{j_2,m-1}\right]^{(1)};x_2,z_2 \bigr) \Bigr\rangle \nonumber \\
& \qquad = \frac{\left(h-1-\frac{k}{2}+j_1\right)\left(h-1-\frac{k}{2}+j_2\right)}{(z_1-z_2)^4} \langle V^1_{h}\left(x_1;z_1\right)V^1_{h}\left(x_2;z_2\right) \rangle \nonumber \\ 
& \qquad \qquad -\frac{\partial_{x_1}\partial_{x_2}}{(z_1-z_2)^2} \langle V^1_{h-1}\left(x_1;z_1\right)V^1_{h-1}\left(x_2;z_2\right) \rangle \ .
\label{eq:resultJ+J+}
\end{align}
The other terms contributing to eq.~\eqref{world-sheet-TT} can be computed in a similar manner, and the explicit results can be found in Appendix~\ref{app:two-point-function}. In order to compute some of the most tedious terms, we have made use of the  Thielemans OPE package \cite{Thielemans:1991uw}, and of the \emph{Virasoro} package developed by Matthew Headrick \cite{Virasoropackage}. 

\subsection{Recovering the dual CFT correlators from the world-sheet}

We can now put the various pieces of the calculation together. Let us first consider the $3$-point functions of eq.~(\ref{ratio-stress-tensor}).

\subsubsection[The $\langle \sigma_w \,  T \,  \sigma_w \rangle $ correlator]{The $\boldsymbol{\langle \sigma_w \,  T \,  \sigma_w \rangle}$ correlator}

Combing eqs.~(\ref{stress-tensor-tot-simplified}) and (\ref{stress-tensor-L-simplified}) with the results of Appendix~\ref{app:formulas-for-additional-modes}, we can express eq.~(\ref{ratio-stress-tensor}) 
in terms of the unknowns $F^i_\ell$.  As explained in  Section~\ref{sec:ward}, the $\mathfrak{sl}(2,\mathds{R})$ Ward identities allow us to determine these unknowns in terms of the corresponding correlators of primary fields
\be 
\left\langle V^w_{h_1}(x_1;z_1) V^1_{h_2}(x_2;z_2) V^w_{h_3}(x_3;z_3) \right\rangle \ , 
\label{eq:three-point-primaries}
\ee
with in general shifted values of $h_i$. While the Ward identities do not fix the latter correlators completely, there exists a natural answer \cite{Eberhardt:2019ywk}, namely eq.~(\ref{solution-equation-5.8}), provided that the spins sum up to $\frac{k}{2}$, see eq.~(\ref{j-constraint}). We will in the following work with this solution,\footnote{In the supersymmetric case based on $\mathfrak{psu}(1,1|2)_1$, it was recently shown in \cite{Dei:2020zui} that this is in fact the only solution.} and choose the $j_i$ as 
\be
j_1=\frac{k-2}{2} \ , \qquad j_2=\frac{k-2}{2} \ , \qquad  j_3=1-\frac{k-2}{2} \ ,
\label{eq:choice-ji}
\ee
so that eq.~(\ref{j-constraint}) is satisfied. We have checked that the result is unchanged if we permute the roles of $j_1$, $j_2$ and $j_3$ in \eqref{eq:choice-ji}.

We can therefore determine the correlators in (\ref{ratio-stress-tensor}) completely, except that the solution for the unknowns $F^i_\ell$ is not known in closed form, but needs to be worked out case by case. We have performed this analysis for $w=1,2,\ldots,6$ (with the help of {\tt Mathematica}), and we find
\begin{align}
\langle V^{w}_{h_3}\left(\infty;\infty\right)  V \left(T;x,z \right)\,  V^{w}_{h_1}\left(0;0\right) \rangle
= C(j_1,j_2,j_3)\, \frac{6k (w^2-1)}{24 w^2}  \frac{1}{x^2}\, \frac{1}{z} \ .
\label{ratio-stress-tensor-total-result}
\end{align}
For a $3$-point function on the world-sheet, the integral over the world-sheet moduli space is trivial since we can use the M\"obius symmetry to set $z=1$. Dividing by the correlator where we replace $T$ with the vacuum vertex operator $V^1_0(x;z)$ removes the factor of $C(j_1,j_2,j_3)$, and we thus reproduce the dual CFT answer eq.~\eqref{sTsans} with $c_{\text{seed}}=6k$.

\subsubsection[The $\langle \sigma_w \,  T^\text{L} \,  \sigma_w \rangle $ correlator]{The $\boldsymbol{\langle \sigma_w \,  T^\text{L} \,  \sigma_w \rangle }$ correlator}

The analysis of the correlator where we replace $T$ by $T^{\text{L}}$ is similar. The only difference is that now not just the unknowns $F_i^\ell$ appear at an intermediate stage, but also  $G_i(\ell,m)$ and $M_{i,p}(\ell,m)$, see e.g.\ eq.~\eqref{eq:J3J3}. Again, we cannot solve for them analytically, but we have worked them out case by case, using the relations of Section~\ref{ward-identities-des}, as well as {\tt Mathematica}; for $w=2,3$ we have confirmed that\footnote{Since the system of relations of Section~\ref{ward-identities-des} is overcomplete, it is a non-trivial consistency check that a solution exists at all.} 
\begin{align}
\langle V^{w}_{h_3}\left(\infty;\infty\right)  V \left(T^{\text{L}};x,z \right)\,  V^{w}_{h_1}\left(0;0\right) \rangle
=  C(j_1,j_2,j_3)\, \frac{c^{\text L}_{\text{seed}} \, (w^2-1)}{24 \, w^2}\, \frac{1}{x^2}  \, \frac{1}{z}\ , 
\label{ratio-stress-tensor-L-result}
\end{align}
where $c^{\text L}_{\text{seed}}$ is given in eq.~(\ref{cseed}). Thus we reproduce again (\ref{sTsans}).

\subsubsection[The $\langle T  T  \rangle$ correlator]{ The $\boldsymbol{\langle T  T  \rangle }$ correlator }\label{sec:4.3.3}

Given the results of Section \ref{sec:wrapping-of-modes} and Appendix \ref{app:two-point-function}, it is also straightforward to compute the correlator \eqref{world-sheet-TT}. For the choice of spins
\be \label{jchoice}
j_1 = 1-j_2 = \frac{k-2}{2} \ , 
\ee
the analogue of \eqref{solution-equation-5.8} is simply 
\be 
\langle V^1_{h} \left(x_1;z_1\right) V^1_{h}\left(x_2;z_2\right) \rangle = \frac{C}{(x_1-x_2)^{2h}(z_1-z_2)^{2\Delta}} \ , 
\ee
where $C$ is an arbitrary constant, and $\Delta$ denotes the conformal dimension on the world-sheet. Collecting all the terms in eqs.~\eqref{eq:resultJ+J+}, \eqref{eq:resultJ+J3J+} -- \eqref{eq:resultJ3J3J3J+} and setting $h=2$, we then find 
\be \label{TTres}
\left\langle V \left(T;x_1,\infty)\right)\, V \left(T;x_2,0)\right)  \right\rangle
= C\, \frac{6k}{2} \, \frac{1}{(x_1-x_2)^4}\ .
\ee
Dividing by the corresponding vacuum correlator, i.e.\ removing the overall normalisation $C$, we thus reproduce  the dual CFT expectation, see eq.~(\ref{TT}), where we have again used that 
$c_{\text{seed}}=6k$.

We should mention that if instead of (\ref{jchoice}) we  perform the calculation for the choice $j_1=j_2=\frac{k-2}{2}$ (and $k> 5$, say, to avoid some low level exceptions), the result (\ref{TTres}) picks up an additional factor of $(k-2)$, while for the choice $j_1=j_2=2  - \frac{k}{2}$ (again with $k> 5$, say) the additional factor is $\frac{1}{(k-2)}$. We do not have a detailed understanding of these phenomena, except that these factors are trivial for $k=3$ where the dual Liouville theory does not have a background charge.\footnote{Recall that $k=3$ for the bosonic algebra corresponds to the supersymmetric case with $k_{\rm susy}=1$, which is the example that has been established best.} We also note that in the general case with   background charge, the choice (\ref{jchoice}) is the most natural one, see eq.~(5.43) of \cite{Eberhardt:2019ywk}.  

\subsubsection[The $\langle T^\text{L}  T^\text{L}  \rangle$ correlator]{ The $\boldsymbol{\langle T^\text{L}  T^\text{L}  \rangle }$ correlator }\label{sec:4.3.4}

The computation of the correlator where we replace $T$ by $T^{\text{L}}$ works similarly. Collecting all the terms in eqs.~\eqref{eq:resultJ+J+}, \eqref{eq:resultJ+J3J+} -- \eqref{eq:resultJ3J3J3J3} and setting $h=2$, we find 
\be \label{TTresL}
\left\langle V \left(T^\text{L};x_1,\infty)\right)\, V \left(T^\text{L};x_2,0)\right)  \right\rangle
= C\, \frac{ c^{\text{L}}_{\text{seed}}}{2}  \, \frac{1}{(x_1-x_2)^4}\ ,
\ee
where we have worked with the choice (\ref{jchoice}), and $c^{\text{L}}_{\text{seed}}$ is given by \eqref{cseed}. Again, after dividing by the corresponding vacuum correlator, we reproduce the dual CFT expectation, see eq.~\eqref{TT}. If we replace in (\ref{jchoice}) $j_1\mapsto 2 - \frac{k}{2}$ or $j_2 \mapsto \frac{k-2}{2}$, we pick up some rather complicated rescaling factors, but as above they become trivial (once suitably regularised) for $k\rightarrow 3$. 

\section{Conclusions}\label{sec:concl}

In this paper we have calculated some simple world-sheet correlators that correspond to stress-energy correlators in the dual CFT, see in particular the $2$-point functions of eq.~(\ref{TTres}) and the $3$-point functions of eq.~(\ref{ratio-stress-tensor-total-result}). While the calculations on the world-sheet are quite complicated --- in general the more general Ward identities of Section~\ref{ward-identities-des} had to be used in order to evaluate them --- the main motivation for the paper was to show that (a) one can in principle determine them with these generalised techniques; and (b) they reproduce the expected conformal dimensions of, e.g., the twisted sector ground states. In the process we also clarified which symmetric orbifold field is actually dual to the `stress-energy' tensor vertex operator on the world-sheet, see the discussion in Section~\ref{sec:2.1}. 

It would be interesting to generalise the analysis of this paper to the supersymmetric theory, in particular using the free field realisation of $\mathfrak{psu}(1,1|2)_1$ that was recently employed in \cite{Dei:2020zui}. It would also be interesting to use similar techniques to rederive the symmetry algebra of the dual CFT from the world-sheet; this would complement the analysis of \cite{Eberhardt:2019qcl,Dei:2019osr,Dei:2019iym}, where this was done algebraically. As we noted in Section~\ref{ward-identities-des}, the Ward identities for the descendant states lead to an overconstrained system; it would be interesting to see whether this gives additional constraints on the solutions (and maybe even shows that eq.~(\ref{solution-equation-5.8}) is in fact the only solution). Recently, the large twist limit of these correlators was shown to have a simple solution  \cite{Gaberdiel:2020ycd}; it would be interesting to analyse whether (and if so how) the Ward identities of the correlators also reflect this simplification.

\acknowledgments 
This work is largely based on the Master thesis of one of us (H.B.). The work of A.D.\ and M.R.G.\ is supported by the Swiss National Science Foundation via a personal grant, as well as through the NCCR SwissMAP. We thank Lorenz Eberhardt for useful conversations, for his comments on a draft version of this paper, and for collaboration on related topics. 

\appendix

\section{Coefficients for the total and the Liouville stress-energy tensor}
\label{app:stress-tensor-coefficients}

The coefficients entering eq.~\eqref{stress-tensor-tot} read \cite{Dei:2019osr}
\begin{subequations}
\begin{align}
c_{++} &= \frac{3 \left(2 j^2+j (4 k-6)+k^2-k (m+3)+4\right)}{2 (j+k-2) (2 j+k-2) (2 j+k-1)} \ , \\
c_{3+} &= -\frac{3 \left(4 j^2+6 j (k-2)+k (k-2 m-5)+8\right)}{(j+k-2) (2 j+k-2) (2 j+k-1)} \ , \\
c_{-+} &= \frac{6 (j-1) (j+k-2)-3 k m}{(j+k-2) (2 j+k-2) (2 j+k-1)} \ , \\
c_{-3} &= \frac{3 \left(-4 j^2-2 j (k-6)+k (k+2 m+3)-8\right)}{(j+k-2) (2 j+k-2) (2 j+k-1)} \ , \\
c_{--} &= \frac{6 (j-3) j-3 k (k+m+1)+12}{2 (j+k-2) (2 j+k-2) (2 j+k-1)} \ , \\
c_{33} &= \frac{12 (j-1) (j+k-2)-6 k m}{(j+k-2) (2 j+k-2) (2 j+k-1)} \ , \\
c_{3} &= \frac{3 k m-3 (j-1) \left(4 j^2+4 j (k-2)+(k-3) k\right)}{(j+k-2) (2 j+k-2) (2 j+k-1)} \ , \\
c_{+} &= \frac{3 \left(8 j^3+12 j^2 (k-2)+2 j (k-2) (3 k-4)+k ((k-5) k-2 m+8)\right)}{2 (j+k-2) (2 j+k-2) (2 j+k-1)} \ , \\
c_{-} &= \frac{8 j^3+4 j^2 (k-4)-2 j (k (k+2)+4)-k \left(k^2+k+6 m+2\right)+16}{2 (j+k-2) (2 j+k-2) (2 j+k-1)}  \ ,
\label{coefftot}
\end{align}
\end{subequations}
while those entering eq.~\eqref{stress-tensor-L} are\ \cite{Dei:2019osr}
\begin{subequations}
\begin{align}
\text{C} &= \frac{1}{4(k-2)(j+k-2)(2 j+k-2)(2 j+k-1)} \ , \\
c_{++}^{\textnormal{L}} &= \text{C} \Bigl(12 j^2 (k-2)+j \left(24 k^2-61 k+20\right)+6 k^3   \nonumber \\  
& \qquad \quad - 6 k^2 (m+5)+k (83-11 m)+52 m-100 \Bigr) \ ,   \\
c_{3+}^{\textnormal{L}} &=-2 \text{C} \Bigl(24 j^2 (k-2)+36 j (k-2)^2+6 k^3  \nonumber \\
&\qquad \quad - k^2 (12 m+65)+k (229-22 m)+4 (26 m-63)\Bigr) \ ,   \\
c_{-+}^{\textnormal{L}} &=2 \text{C} \Bigl(12 j^2 (k-2)+j \left(12 k^2-83 k+124\right)  \nonumber \\  
& \qquad \quad - k^2 (6 m+35)+k (146-11 m)+52 m-152 \Bigr) \ ,   \\
c_{-3}^{\textnormal{L}} &=-2 \text{C} \Bigl(24 j^2 (k-2)+4 j \left(3 k^2-47 k+88\right)-6 k^3  \nonumber \\  
&\qquad \quad - 3 k^2 (4 m+25)+k (355-22 m)+4 (26 m-89)\Bigr) \ ,   \\
c_{--}^{\textnormal{L}} &=\text{C} \Bigl(12 j^2 (k-2)-3 j (35 k-76)-6 k^3-2 k^2 (3 m+20)  \nonumber \\  
&\qquad \quad - 11 k (m-19)+52 m-204\Bigr) \ ,   \\
c_{33}^{\textnormal{L}} &=4 \text{C} \Bigl(12 j^2 (k-2)+j \left(12 k^2-83 k+124\right)-k^2 (6 m+35)  \nonumber  \\  
& \qquad \quad + k (146-11 m)+52 m-152 \Bigr) \ ,   \\
c_{3}^{\textnormal{L}} &=-2 \text{C} \Bigl(24 j^3 (k-2)+4 j^2 \left(6 k^2-53 k+88\right) \nonumber \\ 
& \qquad \qquad + j \left(6 k^3-146 k^2+501 k-460\right)-29 k^3 \nonumber \\   
& \qquad \qquad +k^2 (151-6 m) - k (11 m+215)+52 (m+1)\Bigr) \ ,  \\
c_{+}^{\textnormal{L}} &=\text{C} \Bigl(48 j^3 (k-2)+4 j^2 \left(18 k^2-95 k+124 \right) \nonumber \\ 
& \qquad \quad + 4 j \left(9 k^3-71 k^2+182 k-152\right)+6 k^4 \nonumber \\ 
& \qquad \quad -65 k^3 +k^2 (229-12 m) - 2 k (11 m+126)+104 m \Bigr) \ , \\
c_{-}^{\textnormal{L}} &=\text{C} \Bigl(16 j^3 (k-2)+4 j^2 \left(2 k^2-35 k+68\right) \nonumber\\ 
& \qquad \quad - 4 j \left(k^3+23 k^2-75 k+44\right)-2 k^4-21 k^3 \nonumber \\ 
& \qquad \quad +k^2 (75-12 m) + k (34-22 m)+8 (13 m-21)\Bigr) \ .
\label{coeffL}
\end{align}
\end{subequations}
\smallskip 

\noindent The coefficients $c_1, \dots,  c_6$ in eq.~\eqref{stress-tensor-tot-simplified} are
\be 
c_1= -1 \ , \qquad c_5=2 \ , \qquad c_2=c_3=c_4=c_6= 0 \ , 
\ee
for $j=\frac{k-2}{2}$ and
\begin{subequations}
\begin{align}
c_1 & = -1  \ , & 
c_2 & =-\frac{2 \left(k^2-5 k+6\right)}{(k-5) (k-4)} \ , \\
c_3 & = \frac{2 \left(k^3-9 k^2+32 k-42\right)}{(k-5) (k-4)} \ , & 
c_4 & = \frac{(-k^3+11 k^2-42 k+54)}{(k-5) (k-4)}\ , \\ 
c_5 & = 2 \ , &
c_6 & = -\frac{2 \left(k^2-5 k+6\right)}{(k-5) (k-4)} \ ,
\end{align}
\end{subequations}
for $j=1-\frac{k-2}{2}$. Similarly, the coefficients $c_1^{\textnormal{L}}, \dots,  c_6^{\textnormal{L}}$ in eq.~\eqref{stress-tensor-L-simplified} are
\begin{subequations}
\begin{align}
c_1^{\textnormal{L}} & = \frac{(-12 k^2+65 k-88)}{6 (k-2) (2 k-3)}  \ , & 
c_2^{\textnormal{L}} & = \frac{(52-23 k) }{(24 k^2-84 k+72)}  \ , \\
c_3^{\textnormal{L}} & = \frac{(23 k-52)}{4 \left(2 k^2-7 k+6\right)} \ , & 
c_4^{\textnormal{L}} & = \frac{(52-23 k)}{24 \left(2 k^2-7 k+6\right)} \ , \\ 
c_5^{\textnormal{L}} & = \frac{(25 k^2-139 k+196)}{(24 k^2-84 k+72)} \ , &
c_6^{\textnormal{L}} & = \frac{(k+16)}{36-24 k} \ ,
\end{align}
\end{subequations}
for $j=\frac{k-2}{2}$ and
\begin{subequations}
\begin{align}
c_1^{\textnormal{L}} & = \frac{(-6 k^2+35 k-52)}{6 (k-2) k} \ , \\ 
c_2^{\textnormal{L}} & = \frac{(-12 k^4-8 k^3+591 k^2-2076 k+2080)}{6 (k-5) (k-4) (k-2) k}  \ , \\
c_3^{\textnormal{L}} & = \frac{(12 k^5-109 k^4+548 k^3-1916 k^2+3844 k-3120)}{6 (k-5) (k-4) (k-2) k} \ , \\
c_4^{\textnormal{L}} & = \frac{(-12 k^4+109 k^3-314 k^2+106 k+520)}{12 (k-5) (k-4) k} \ , \\ 
c_5^{\textnormal{L}} & = \frac{(6 k^2-35 k+52)}{3 (k-2) k} \ , \\
c_6^{\textnormal{L}} & = \frac{(-18 k^3+45 k^2+194 k-520)}{6 (k-5) (k-4) k} \ ,
\end{align}
\end{subequations}
for $j=1-\frac{k-2}{2}$.

\section{Wrapping of modes for the three-point function}
\label{app:formulas-for-additional-modes}

The various terms of the form \eqref{3pt-function-general} read
\begin{align}
& \Bigl\langle V\bigl(\left[J_{-1}^{3}J_{-1}^{3}\ket{j_i,m_i}\right]^{(1)}; x_i,z_i\bigr)\, \prod_{l\neq i}V_{h_l}^{w_l}\Bigr\rangle \nonumber \\ 
& \qquad =  \sum_{j\neq i} \Biggl[ \frac{h_j + (x_j-x_i)\partial_{x_j}}{(z_j-z_i)}\Biggl(\frac{h_j + (x_j-x_i)\partial_{x_j}}{(z_j-z_i)} \langle \prod_{l}V^{w_l}_{h_l} \rangle   \nonumber \\ 
& \hspace{60pt} +  (x_j-x_i) \sum_{\ell=1}^{w_j} \frac{(-1)^\ell}{(z_j-z_i)^{1+\ell}} \bigl\langle [J_\ell^+V^{w_j}_{h_j}]\, \prod_{l\neq j} V^{w_l}_{h_l} \bigr\rangle \Biggr)  \nonumber \\
& \hspace{60pt} + (x_j-x_i)\sum_{r=1}^{w_j} \frac{(-1)^r}{(z_j-z_i)^{1+r}} \Biggl(\frac{h_j+1+(x_j-x_i)\partial_{x_j}}{(z_j-z_i)} \bigl\langle  [J^+_rV^{w_j}_{h_j}]\, \prod_{l\neq j} V^{w_l}_{h_l} \bigr\rangle   \nonumber \\ 
& \hspace{60pt} +  \sum_{\ell=1}^{w_j-r} \frac{(-1)^\ell}{(z_j-z_i)^{\ell+1}} \bigl\langle [J^+_{\ell+r}V^{w_j}_{h_j}]\, \prod_{l\neq j} V^{w_l}_{h_l} \bigr\rangle   \nonumber \\ 
& \hspace{60pt}  +  (x_j-x_i)\sum_{\ell=1}^{w_j} \frac{(-1)^\ell}{(z_j-z_i)^{\ell+1}} \bigl\langle [J_r^+J_\ell^+V^{w_j}_{h_j}]\, \prod_{l\neq j} V^{w_l}_{h_l} \bigr\rangle \Biggr) \Biggr] \nonumber \\
& \qquad \quad +\sum_{p \neq j,i} \Biggl[ \frac{h_j + (x_j-x_i)\partial_{x_j}}{(z_j-z_i)}\Biggl(\frac{h_p + (x_p-x_i)\partial_{x_p}}{(z_p-z_i)} \langle \prod_{l}V^{w_l}_{h_l} \rangle   \nonumber \\
& \hspace{60pt} + (x_p-x_i) \sum_{\ell=1}^{w_p} \frac{(-1)^\ell}{(z_p-z_i)^{\ell+1}} \bigl\langle [J_\ell^+V^{w_p}_{h_p}]\, \prod_{l\neq p} V^{w_l}_{h_l} \bigr\rangle \Biggr)  \nonumber \\
& \hspace{60pt} +  (x_j-x_i)\sum_{r=1}^{w_j} \frac{(-1)^r}{(z_j-z_i)^{1+r}} \Biggl( \frac{h_p+(x_p-x_i)\partial_{x_p}}{(z_p-z_i)} \bigl\langle  [J^+_rV^{w_j}_{h_j}]\, \prod_{l\neq j} V^{w_l}_{h_l} \bigr\rangle  \nonumber \\ 
& \hspace{60pt} + (x_p-x_i)\sum_{\ell=1}^{w_p} \frac{(-1)^\ell}{(z_p-z_i)^{\ell+1}} \bigl\langle [J_r^+V^{w_j}_{h_j}]\, [J_\ell^+V^{w_p}_{h_p}]\, \prod_{l\neq p,j} V^{w_l}_{h_l} \bigr\rangle \Biggr) \Biggr] \ , \label{eq:J3J3} \\
& \ \nonumber \\
& \Bigl\langle V\bigl(\left[J_{-1}^{+}J_{-1}^{+}\ket{j_i, m_i-2}\right]^{(1)}; x_i,z_i \bigr)\prod_{l\neq i}V_{h_l}^{w_l} \Bigr\rangle = \partial^2_{x_i} \Bigl\langle V_{h_i-2}^{1}\prod_{l\neq i}V_{h_l}^{w_l} \Bigr\rangle \ , \label{eq:J+J+} \\
& \ \nonumber \\ 
&\Bigl\langle V\bigl(\left[J_{-1}^{-}J^+_{-1}\ket{j_i, m_i}\right]^{(1)};x_i,z_i\bigr) \prod_{l\neq i}V_{h_l}^{w_l} \Bigr\rangle \nonumber \\ 
& \qquad = -\sum_{j\neq i} \Biggl[\frac{2h_j(x_j-x_i)+(x_j-x_i)^2 \partial_{x_j}}{(z_j-z_i)^2} \partial_{x_i}\langle \prod_{l}V^{w_l}_{h_l}\rangle \nonumber \\ 
& \hspace{60pt} + (x_j-x_i)^2\sum_{\ell=1}^{w_j}\frac{(-1)^\ell(\ell+1)}{(z_j-z_i)^{\ell+2}} \partial_{x_i} \bigl\langle[J^+_\ell V^{w_j}_{h_j}]\, \prod_{l\neq j}V^{w_l}_{h_l}\bigr\rangle \Biggr] \ , \label{eq:J-J+} \\
& \  \nonumber \\
& \Bigl\langle V\bigl(\left[J_{-1}^{3}J^+_{-1}\ket{j_i, m_i-1}\right]^{(1)}; x_i,z_i\bigr)\prod_{l\neq i}V_{h_l}^{w_l} \Bigr\rangle \nonumber \\ 
& \qquad = -\sum_{j\neq i} \Biggl[\frac{h_j+(x_j-x_i)\partial_{x_j}}{(z_j-z_i)} \partial_{x_i}\langle V_{h_i-1}^{1}\prod_{l\neq i}V^{w_l}_{h_l}\rangle \nonumber \\ 
& \hspace{60pt}+ (x_j-x_i)\sum_{\ell=1}^{w_j}\frac{(-1)^\ell}{(z_j-z_i)^{\ell+1}} \partial_{x_i}\langle [J^+_\ell V^{w_j}_{h_j}]\, V^{1}_{h_i-1}\prod_{l\neq j,i}V^{w_l}_{h_l}\rangle\Biggr] \ , \label{eq:J3J+} \\
& \ \nonumber \\
&\Bigl\langle V\bigl(\left[J_{-2}^{+}\ket{j_i, m_i-1}\right]^{(1)}; x_i,z_i\bigr)\prod_{l\neq i}V_{h_l}^{w_l} \Bigr\rangle \nonumber \\ 
& \qquad = -\sum_{j\neq i} \Biggl[\frac{\partial_{x_j}}{(z_j-z_i)} \langle V^{1}_{h_i-1} \prod_{l \neq i}V^{w_l}_{h_l}\rangle \nonumber \\ 
& \hspace{60pt} + \sum_{\ell=1}^{w_j}\frac{(-1)^\ell}{(z_j-z_i)^{\ell+1}} \bigl\langle [J^+_\ell V^{w_j}_{h_j}]\, V^{1}_{h_i-1}\prod_{l\neq j,i}V^{w_l}_{h_l} \bigr\rangle \Biggr] \ .
\label{eq:J+}
\end{align}

\section{Wrapping of modes for the two-point function}
\label{app:two-point-function}

In this appendix we collect the explicit results for the computation of the correlator
\be 
\left\langle V \left(T;x_1,z_1 \right) V \left(T;x_2,z_2 \right) \right\rangle \ .
\ee
In Section \ref{sec:computations} we have already computed the term \eqref{eq:resultJ+J+}. Following a similar procedure, for the other terms appearing in eq.~\eqref{stress-tensor-tot-simplified} we find 
\begin{align}
& \Bigl\langle V\bigl(J^+_{-1}\left[\ket{j_1,m-1}\right]^{(1)};x_1,z_1\bigr)\, 
V\bigl(J^3_{-1}J^+_{0}\left[\ket{j_2,m-1}\right]^{(1)};x_2,z_2\bigr) \Bigr\rangle \nonumber \\
& \qquad = \frac{(x_1-x_2)\left(h-1-\frac{k}{2}+j_1\right)\left(h-1-\frac{k}{2}+j_2\right)}{(z_1-z_2)^4} \partial_{x_2}  \langle V^1_{h}\left(x_1;z_1\right)V^1_{h}\left(x_2;z_2\right) \rangle \nonumber \\ 
& \qquad \quad + \left(\partial_{x_1} -h\, \partial_{x_2}-(x_1-x_2)\partial_{x_1} \partial_{x_2} \right)  \partial_{x_2} \frac{\langle V^1_{h-1} \left(x_1;z_1\right)V^1_{h-1}\left(x_2;z_2\right) \rangle}{(z_1-z_2)^2} \ , \label{eq:resultJ+J3J+} \\[4pt]
& \Bigl\langle V\bigl(J^3_{-1}J^+_{0} \left[\ket{j_1,m-1}\right]^{(1)};x_1,z_1\bigr)\, 
V\bigl(J^3_{-1}J^+_{0}\left[\ket{j_2,m-1}\right]^{(1)};x_2,z_2\bigr)  \Bigr\rangle \nonumber \\
& \qquad = \left(1-(x_1-x_2)\partial_{x_2}+(x_1-x_2)\partial_{x_1} - (x_1-x_2)^2 \partial_{x_1} \partial_{x_2} \right) \,  \nonumber \\
& \qquad \qquad \times \frac{\left(h-1-\frac{k}{2}+j_1 \right)\left(h-1-\frac{k}{2}+j_2\right)\langle V^1_{h} \left(x_1;z_1\right) V^1_{h}\left(x_2;z_2\right) \rangle}{(z_1-z_2)^4} \nonumber \\
& \qquad \quad + \Bigl( (x_1-x_2)^2 \partial_{x_1} \partial_{x_2} + h(x_1-x_2) \partial_{x_2} -h(x_1-x_2) \partial_{x_1}  - (x_1-x_2) \partial_{x_1} \nonumber  \\
& \hspace{90pt}  + (x_1-x_2) \partial_{x_2} - h^2-\tfrac{k}{2} \Bigr) \frac{\partial_{x_1} \partial_{x_2} \langle V^1_{h-1}\left(x_1;z_1\right)V^1_{h-1}\left(x_2;z_2\right) \rangle}{(z_1-z_2)^2} \ , 
\label{eq:resultJ3J+J3J+} \\[4pt]
& \Bigl\langle V\bigl(J^+_{-1}\left[\ket{j_1,m-1}\right]^{(1)};x_1,z_1\bigr)\, 
V\bigl(J^3_{-2}\left[\ket{j_2,m}\right]^{(1)};x_2,z_2\bigr) \Bigr\rangle \nonumber \\
& \qquad = \frac{\left(h-1-\frac{k}{2}+j_1\right)}{(z_1-z_2)^4} \left( -3 +2 (x_1 - x_2) \partial_{x_2} \right)  \langle V^1_{h}\left(x_1;z_1\right)V^1_{h}\left(x_2;z_2\right) \rangle \ , 
\label{eq:resultJ+J3} \\[4pt]
& \Bigl\langle V\bigl(J^+_{-1}\left[\ket{j_1,m-1}\right]^{(1)};x_1,z_1\bigr)\, 
V\bigl(J^-_{-2}J^+_0\left[\ket{j_2,m}\right]^{(1)};x_2,z_2\bigr) \Bigr\rangle \label{eq:resultJ+J-J+}   \\
& \qquad = \frac{(x_1 -x_2)\left(h-1-\frac{k}{2}+j_1\right)}{(z_1-z_2)^4} \left( -6 +2 (x_1 - x_2) \partial_{x_2} \right) \partial_{x_2} \,  
\langle V^1_{h}\left(x_1;z_1\right)V^1_{h}\left(x_2;z_2\right) \rangle \ , \nonumber
\\[4pt]
& \Bigl\langle V\bigl(J^+_{-1}\left[\ket{j_1,m-1}\right]^{(1)};x_1,z_1\bigr)\, 
V\bigl(J^3_{-1}J^3_{-1}\left[\ket{j_2,m}\right]^{(1)};x_2,z_2\bigr) \Bigr\rangle \nonumber \\
& \qquad = \frac{(x_1 - x_2)^2\left(h-1-\frac{k}{2}+j_1\right)\left(h-\frac{k}{2}+j_1\right)\left(h-\frac{k}{2}+j_2\right)}{(z_1-z_2)^6} \,  
\langle V^1_{h+1}\left(x_1;z_1\right)V^1_{h+1}\left(x_2;z_2\right) \rangle \nonumber \\ 
& \qquad \quad + \frac{\left(h-1-\frac{k}{2}+j_1\right)}{(z_1-z_2)^4}\Bigl( 2h+1-2(h+1)(x_1-x_2) \partial_{x_2}+4(x_1-x_2)\partial_{x_1} \nonumber  \\
& \hspace{90pt}-2(x_1-x_2)^2 \partial_{x_1}\partial_{x_2} \Bigr) \langle V^1_{h}\left(x_1;z_1\right)V^1_{h}\left(x_2;z_2\right) \rangle \ , 
\label{eq:resultJ+J3J3} \\[4pt]
& \Bigl\langle V\bigl(J^-_{-2}J^+_0\left[\ket{j_1,m}\right]^{(1)};x_1,z_1\bigr)\, 
V\bigl(J^3_{-1}J^+_{0}\left[\ket{j_2,m-1}\right]^{(1)};x_2,z_2\bigr) \Bigr\rangle \nonumber \\
& \qquad = \frac{(x_1 - x_2)\left(h-1-\frac{k}{2}+j_2\right)}{(z_1-z_2)^4}\Bigl( -4(h+1)+(2h+5)(x_1-x_2)\partial_{x_2}\nonumber  \\
& \hspace{70pt} -4(x_1-x_2)\partial_{x_1}+2(x_1-x_2)^2 \partial_{x_1}\partial_{x_2} \Bigr)\partial_{x_1} \langle V^1_{h}\left(x_1;z_1\right)V^1_{h}\left(x_2;z_2\right) \rangle \ , 
\label{eq:resultJ-J+J3J+} \\[4pt]
& \Bigl\langle V\bigl(J^3_{-1}J^+_0\left[\ket{j_1,m-1}\right]^{(1)};x_1,z_1\bigr)\, 
V\bigl(J^3_{-2}\left[\ket{j_2,m}\right]^{(1)};x_2,z_2\bigr) \Bigr\rangle \nonumber \\
& \qquad = \frac{\left(h-1-\frac{k}{2}+j_1\right)}{(z_1-z_2)^4}\Bigl( 2h-2(x_1-x_2)\partial_{x_2}+(2h+3)(x_1-x_2)\partial_{x_1}\nonumber  \\
& \hspace{90pt} -2(x_1-x_2)^2 \partial_{x_1} \partial_{x_2} \Bigr) \langle V^1_{h}\left(x_1;z_1\right)V^1_{h}\left(x_2;z_2\right) \rangle \ , 
\label{eq:resultJ3J+J3} \\[4pt]
& \Bigl\langle V\bigl(J^3_{-1}J^3_{-1}\left[\ket{j_1,m}\right]^{(1)};x_1,z_1\bigr)\, 
V\bigl(J^3_{-1}J^+_0\left[\ket{j_2,m-1}\right]^{(1)};x_2,z_2\bigr) \Bigr\rangle \nonumber \\
& \qquad = \frac{(x_1 - x_2)^2\left(h-1-\frac{k}{2}+j_2\right)\left(h-\frac{k}{2}+j_1\right)\left(h-\frac{k}{2}+j_2\right)}{(z_1-z_2)^6} \,   \nonumber \\
& \qquad \qquad \quad \times \Bigl(-2+(x_1-x_2)\partial_{x_2} \Bigr) \langle V^1_{h+1}\left(x_1;z_1\right)V^1_{h+1}\left(x_2;z_2\right) \rangle \nonumber \\ 
& \qquad \quad + \frac{\left(h-1-\frac{k}{2}+j_2\right)}{(z_1-z_2)^4}\Bigl( -2h^2-k+(5+6h+2h^2+k)(x_1-x_2) \partial_{x_2}  \nonumber  \\
& \hspace{75pt} -2(h+1)(x_1-x_2)\partial_{x_1}-2(h+2)(x_1-x_2)^2 \partial_{x_2}^2 \label{eq:resultJ3J3J3J+}   \\
& \hspace{75pt} +2(h+3)(x_1-x_2)^2 \partial_{x_1}\partial_{x_2}-2(x_1-x_2)^3 \partial_{x_1} \partial_{x_2}^2 \Bigr) \, 
\langle V^1_{h}\left(x_1;z_1\right)V^1_{h}\left(x_2;z_2\right) \rangle \ . \nonumber
\end{align}
\smallskip

\noindent Similarly, in order to compute the correlator
\be 
\left\langle V \left(T^\text{L};x_1,z_1 \right) V \left(T^\text{L};x_2,z_2 \right) \right\rangle \ ,
\ee
we need the following additional terms
\begin{align}
& \Bigl\langle V\bigl(L^X_{-2}\left[\ket{j_1,m}\right]^{(1)};x_1,z_1\bigr)\, 
V\bigl(L^X_{-2}\left[\ket{j_2,m}\right]^{(1)};x_2,z_2\bigr) \Bigr\rangle \nonumber \\
& \qquad = \frac{1}{(z_1-z_2)^4} \left(13-\tfrac{3k}{2(k-2)} \right) \langle V^1_{h}\left(x_1;z_1\right)V^1_{h}\left(x_2;z_2\right) \rangle \ , \label{eq:resultLXLX} \\[4pt]
& \Bigl\langle V\bigl(J^3_{-2}\left[\ket{j_1,m}\right]^{(1)};x_1,z_1\bigr)\, 
V\bigl(J^3_{-2}\left[\ket{j_2,m-1}\right]^{(1)};x_2,z_2\bigr) \Bigr\rangle \nonumber \\
& \qquad = \frac{4(x_1-x_2)^2\left(h-\frac{k}{2}+j_1\right)\left(h-\frac{k}{2}+j_2\right)}{(z_1-z_2)^6}  \langle V^1_{h+1}\left(x_1;z_1\right)V^1_{h+1}\left(x_2;z_2\right) \rangle \nonumber \\ 
& \qquad \quad +  \frac{1}{(z_1-z_2)^4}\Bigl(h^2+3k+ (3+h)(x_1-x_2)\partial_{x_1}-(3+h)(x_1-x_2)\partial_{x_2} \nonumber  \\
& \hspace{90pt} -(x_1-x_2)^2\partial_{x_1}\partial_{x_2} \Bigr)\langle V^1_{h} \left(x_1;z_1\right)V^1_{h}\left(x_2;z_2\right) \rangle \ , \label{eq:resultJ3J3} \\[4pt]
& \Bigl\langle V\bigl(J^3_{-2}\left[\ket{j_1,m}\right]^{(1)};x_1,z_1\bigr)\, 
V\bigl(J^-_{-2}J^+_0\left[\ket{j_2,m}\right]^{(1)};x_2,z_2\bigr) \Bigr\rangle \nonumber \\
& \qquad = \frac{4(x_1-x_2)^2\left(h-\frac{k}{2}+j_1\right)\left(h-\frac{k}{2}+j_2\right)}{(z_1-z_2)^6} \Bigl( -1+(x_1-x_2)\partial_{x_2} \Bigr) \,   \nonumber \\
& \qquad \qquad \quad \times  
\langle V^1_{h+1}\left(x_1;z_1\right)V^1_{h+1}\left(x_2;z_2\right) \rangle  \nonumber \\[2pt] 
& \qquad \quad +  \frac{1}{(z_1-z_2)^4}\Bigl(6h+2(-3-2h+h^2+3k)(x_1-x_2)\partial_{x_2}-2h(x_1-x_2)^2\partial_{x_2}^2 \label{eq:resultJ3J-J+} \\ 
& \hspace{50pt} + (h+4)(x_1 - x_2)^2 \partial_{x_1} \partial_{x_2} - (x_1 - x_2)^3 \partial_{x_1} \partial_{x_2}^2 \Bigr) \,  
\langle V^1_{h} \left(x_1;z_1\right)V^1_{h}\left(x_2;z_2\right) \rangle \ ,  \nonumber \\[4pt]
& \Bigl\langle V\bigl(J^-_{-2}J^+_0\left[\ket{j_1,m}\right]^{(1)};x_1,z_1\bigr)\, 
V\bigl(J^-_{-2}J^+_0\left[\ket{j_2,m}\right]^{(1)};x_2,z_2\bigr) \Bigr \rangle \nonumber \\
& \qquad = \frac{4(x_1-x_2)^2\left(h-\frac{k}{2}+j_1\right)\left(h-\frac{k}{2}+j_2\right)}{(z_1-z_2)^6} \Bigl( 4 - 2(x_1-x_2) \partial_{x_2} +2(x_1-x_2) \partial_{x_1} \nonumber \\ 
& \hspace{50pt} -(x_1-x_2)^2 \partial_{x_1} \partial_{x_2} \Bigr)  \langle V^1_{h+1}\left(x_1;z_1\right)V^1_{h+1}\left(x_2;z_2\right) \rangle \nonumber \\ 
& \qquad \quad +  \frac{1}{(z_1-z_2)^4}\Bigl(4h^2-4h(h-2)(x_1-x_2)\partial_{x_2}+4h(h-2)(x_1-x_2)\partial_{x_1}  \nonumber  \\
& \hspace{50pt} - 2(-4-2h+2h^2+3k)(x_1 - x_2)^2 \partial_{x_1} \partial_{x_2} + 2h(x_1-x_2)^2 \partial_{x_2}^2 \nonumber  \\
& \hspace{50pt} + 2h (x_1-x_2)^2 \partial_{x_1}^2 -2(h+1)(x_1-x_2)^3 \partial_{x_1}^2 \partial_{x_2} \label{eq:resultJ-J+J-J+} \\
& \hspace{50pt} + 2(h+1)(x_1-x_2)^3 \partial_{x_1} \partial_{x_2}^2 + (x_1 - x_2)^4 \partial_{x_1}^2 \partial_{x_2}^2 \Bigr) \,  
\langle V^1_{h} \left(x_1;z_1\right)V^1_{h}\left(x_2;z_2\right) \rangle \ ,  \nonumber \\[4pt]
& \Bigl\langle V\bigl(J^-_{-2}J^+_0\left[\ket{j_1,m}\right]^{(1)};x_1,z_1\bigr)\, 
V\bigl(J^3_{-1}J^3_{-1}\left[\ket{j_2,m}\right]^{(1)};x_2,z_2\bigr) \Bigr\rangle \nonumber \\
& \qquad = \frac{2(x_1-x_2)^2\left(h-\frac{k}{2}+j_1\right)\left(h-\frac{k}{2}+j_2\right)}{(z_1-z_2)^6} \Bigl(2h+5+(2h+9) (x_1-x_2) \partial_{x_1} \nonumber  \\
& \hspace{90pt} +2(x_1-x_2)^2 \partial_{x_1}^2 \Bigr)  \langle V^1_{h+1}\left(x_1;z_1\right)V^1_{h+1}\left(x_2;z_2\right) \rangle \nonumber \\ 
& \qquad \quad +  \frac{1}{(z_1-z_2)^4}\Bigl(2h(1-2h)+2(h^3-1+2h(k-2)+2k)(x_1-x_2)\partial_{x_1} \nonumber  \\
& \hspace{90pt} - (h^2+6h+8)(x_1 - x_2)^2 \partial_{x_1} \partial_{x_2} -(2h+7)(x_1-x_2)^3 \partial_{x_1}^2 \partial_{x_2}  \label{eq:resultJ-J+J3J3}  \\
& \hspace{90pt} + 2(2h^2+h-1+2k) \,  (x_1-x_2)^2 \partial_{x_1}^2 \nonumber  \\
& \hspace{90pt} + 2h(x_1-x_2)^3 \partial_{x_1}^3 - (x_1 - x_2)^4 \partial_{x_1}^3 \partial_{x_2} \Bigr) \,  
\langle V^1_{h} \left(x_1;z_1\right)V^1_{h}\left(x_2;z_2\right) \rangle \ , \nonumber \\[4pt]
& \Bigl\langle V\bigl(J^3_{-2}\left[\ket{j_1,m}\right]^{(1)};x_1,z_1\bigr)\, 
V\bigl(J^3_{-1}J^3_{-1}\left[\ket{j_2,m}\right]^{(1)};x_2,z_2\bigr) \Bigr\rangle \nonumber \\
& \qquad = \frac{2(x_1-x_2)^2\left(h-\frac{k}{2}+j_1\right)\left(h-\frac{k}{2}+j_2\right)}{(z_1-z_2)^6} \Bigl(-(2h+5) -2 (x_1-x_2) \partial_{x_1} \Bigr) \,   \nonumber \\
& \qquad \qquad \quad \times  \langle V^1_{h+1}\left(x_1;z_1\right)V^1_{h+1}\left(x_2;z_2\right) \rangle \nonumber \\[2pt] 
& \qquad \quad +  \frac{1}{(z_1-z_2)^4}\Bigl(-h(h^2+2k)+(h^2+4h+3)(x_1-x_2)\partial_{x_2}  \label{eq:resultJ3J3J3}  \\
& \hspace{90pt} - (3+3h+2h^2+2k)(x_1 - x_2) \partial_{x_1} + (2h+5) (x_1-x_2)^2 \partial_{x_1} \partial_{x_2} \nonumber  \\
& \hspace{90pt} -(h+2)(x_1-x_2)^2 \partial_{x_1}^2 + (x_1-x_2)^3 \partial_{x_1}^2 \partial_{x_2} \Bigr) \,   
\langle V^1_{h} \left(x_1;z_1\right)V^1_{h}\left(x_2;z_2\right) \rangle \ , \nonumber \\[4pt]
& \Bigl\langle V\bigl(J^3_{-1}J^3_{-1}\left[\ket{j_1,m}\right]^{(1)};x_1,z_1\bigr)\, 
V\bigl(J^3_{-1}J^3_{-1}\left[\ket{j_2,m}\right]^{(1)};x_2,z_2\bigr) \Bigr\rangle \nonumber \\
& \qquad = \frac{(x_1-x_2)^4\left(h-\frac{k}{2}+j_1\right)\left(h-\frac{k}{2}+j_2\right)\left(h+1-\frac{k}{2}+j_1\right)\left(h+1-\frac{k}{2}+j_2\right)}{(z_1-z_2)^8} \,  \nonumber \\
& \qquad \qquad \quad \times \langle V^1_{h+2}\left(x_1;z_1\right)V^1_{h+2}\left(x_2;z_2\right) \rangle \nonumber \\[2pt] 
& \qquad \quad + \frac{(x_1-x_2)^2\left(h-\frac{k}{2}+j_1\right)\left(h-\frac{k}{2}+j_2\right)}{(z_1-z_2)^6} \Bigl(2(10+8h+2h^2+k) \nonumber  \\
& \hspace{90pt} - 4(3+h)(x_1-x_2) \partial_{x_2} + 4(3+h)(x_1-x_2) \partial_{x_1} \nonumber  \\
& \hspace{90pt} - 4 (x_1-x_2)^2 \partial_{x_1} \partial_{x_2} \Bigr)  \langle V^1_{h+1}\left(x_1;z_1\right)V^1_{h+1}\left(x_2;z_2\right) \rangle \nonumber \\ 
& \qquad \quad + \frac{1}{(z_1-z_2)^4}\Bigl(h^4+2h^2k+\tfrac{k^2}{2}-(h+1)(2h^2+3h+3+2k)(x_1-x_2)\partial_{x_2} \nonumber  \\
& \hspace{90pt} +(h+1)(2h^2+3h+3+2k)(x_1-x_2)\partial_{x_1} \nonumber  \\
& \hspace{90pt} - (13+12h+4h^2+2k) (x_1-x_2)^2 \partial_{x_1} \partial_{x_2} \nonumber  \\
& \hspace{90pt} +(h+2)^2(x_1-x_2)^2 \partial_{x_1}^2 +(h+2)^2(x_1-x_2)^2 \partial_{x_2}^2 \nonumber  \\
& \hspace{90pt} - (2h+5)(x_1-x_2)^3 \partial_{x_1}^2 \partial_{x_2} + (2h+5)(x_1-x_2)^3 \partial_{x_1} \partial_{x_2}^2 \nonumber  \\
& \hspace{90pt} + (x_1-x_2)^4 \partial_{x_1}^2 \partial_{x_2}^2 \Bigr)\langle V^1_{h} \left(x_1;z_1\right)V^1_{h}\left(x_2;z_2\right) \rangle \ . \label{eq:resultJ3J3J3J3} 
\end{align}
Of course, given the explicit result for a term of the form \eqref{2pt-function-general}, by exchanging the roles of $j_1, x_1, z_1$ with $j_2, x_2, z_2$, respectively, one immediately obtains the term where $\ket{\phi}_1$ and $\ket{\phi}_2$ have been exchanged.

\end{document}